\documentclass[sigconf]{acmart}

\usepackage{tikz}
\usepackage{pgfplots}
\usetikzlibrary{arrows,arrows.meta,external,colorbrewer,decorations.markings,decorations.pathmorphing,decorations.pathreplacing,external,fit,patterns,shapes,positioning}
\usepgfplotslibrary{groupplots,colormaps,colorbrewer,statistics,fillbetween}
\usepackage[outline]{contour}
\usepackage{pgfplotstable}
\pgfplotsset{compat=1.15}
\usepackage{tcolorbox}
\usepackage[T1]{fontenc}
\usepackage{pgfpages}
\tikzexternaldisable

\usepackage{tabularx}
\usepackage{enumitem}
\usepackage{adjustbox}
\usepackage{makecell}
\usepackage{subfig}
\usepackage{multirow}
\usepackage{hyperref}
\usepackage{booktabs}
\usepackage{balance}
\usepackage{mathtools}
\usepackage{cases}
\usepackage[absolute,showboxes]{textpos}

\usepackage{pifont}
\newcommand{\xmark}{\ding{56}}%

\usepackage{xspace}
\newcommand{\cf}{\textit{cf.}~}
\newcommand{\etal}{\textit{et al.}}
\newcommand{\eg}{\textit{e.g.,}~}
\newcommand{\ie}{\textit{i.e.,}~}
\newcommand{\etc}{\textit{etc.}~}
\newcommand{\one}{({\em i})\xspace}
\newcommand{\two}{({\em ii})\xspace}
\newcommand{\three}{({\em iii})\xspace}
\newcommand{\four}{({\em iv})\xspace}

\let\orgautoref\autoref
\renewcommand{\autoref}
{\def\sectionautorefname{Section}%
\def\subsectionautorefname{Section}%
\def\subsubsectionautorefname{Section}%
\orgautoref}

\makeatletter
\renewcommand{\paragraph}[1]{\vspace*{0.03in}\noindent{\bf #1.}\hspace{0.25ex \@plus1ex \@minus.2ex}}
\newcommand{\paragraphS}[1]{\vspace*{0.03in}\noindent{\bf #1}\hspace{0.25ex \@plus1ex \@minus.2ex}}
\makeatother

\AtBeginDocument{%
}

\setcopyright{acmcopyright}

\acmYear{2022}\copyrightyear{2022}
\setcopyright{acmcopyright}
\acmConference[ICN '22]{9th ACM Conference on Information-Centric Networking}{September 19--21, 2022}{Osaka, Japan}
\acmBooktitle{9th ACM Conference on Information-Centric Networking (ICN '22), September 19--21, 2022, Osaka, Japan}
\acmPrice{15.00}
\acmDOI{10.1145/3517212.3558081}
\acmISBN{978-1-4503-9257-0/22/09}

\begin{document}

\title{Delay-Tolerant ICN and Its Application to LoRa}

\setlength{\TPHorizModule}{.7\textwidth}
\setlength{\TPVertModule}{\paperheight}
\TPMargin{5pt}
\begin{textblock}{1}(.3,0.03)
  \noindent
  \footnotesize
  If you cite this paper, please use the ICN reference:
  P. Kietzmann, J. Alamos, D. Kutscher, T. C. Schmidt, M. W\"ahlisch.\\
  Delay-Tolerant ICN and Its Application to LoRa.
  In \emph{Proc. of ACM ICN}, ACM, 2022.
\end{textblock}

\author{Peter Kietzmann}
\affiliation{%
  \institution{HAW Hamburg}
  \country{Germany}
}
\email{peter.kietzmann@haw-hamburg.de}

\author{Jos{\'e} Alamos}
\affiliation{%
  \institution{HAW Hamburg}
  \country{Germany}
}
\email{jose.alamos@haw-hamburg.de}

\author{Dirk Kutscher}
\affiliation{%
  \institution{Hong Kong University of Science and}
  \country{Technology, China}
}
\email{dku@ust.hk}

\author{Thomas C. Schmidt}
\affiliation{%
  \institution{HAW Hamburg}
  \country{Germany}
}
\email{t.schmidt@haw-hamburg.de}

\author{Matthias W{\"a}hlisch}
\affiliation{%
  \institution{Freie Universit\"at Berlin}
  \country{Germany}
  }
\email{m.waehlisch@fu-berlin.de}

\renewcommand{\shortauthors}{Kietzmann, et al.}

\begin{abstract}
Connecting long-range wireless networks to the Internet imposes challenges due to vastly
longer round-trip-times (RTTs).
In this paper, we present an ICN protocol framework that enables robust and
efficient delay-tolerant communication to edge networks.
Our approach provides ICN-idiomatic
communication between networks with vastly different RTTs.
We applied this framework to LoRa, enabling
end-to-end consumer-to-LoRa-producer interaction over an ICN-Internet
and asynchronous data production in the LoRa edge.  Instead
of using LoRaWAN, we implemented an IEEE 802.15.4e DSME MAC layer on
top of the LoRa PHY and ICN protocol mechanisms in RIOT OS.
Executed on off-the-shelf IoT hardware, we provide a comparative evaluation for basic NDN-style ICN~\cite{zabjc-ndn-14}, RICE~\cite{khokp-rrmii-18}-like pulling, and reflexive forwarding~\cite{draft-oran-icnrg-reflexive-forwarding}. This is the first practical evaluation of ICN over LoRa using a reliable MAC.
Our results show that periodic polling in NDN works inefficiently when facing long and differing RTTs.
RICE reduces polling overhead and exploits gateway knowledge, without violating ICN principles.
Reflexive forwarding reflects sporadic data generation naturally.
Combined with a local data push, it operates efficiently and enables lifetimes of >1\,year for battery powered LoRa-ICN nodes.

\end{abstract}



\maketitle

\section{Introduction}\label{sec:intro}

\begin{figure}[]
  \centering
  \includegraphics[width=1\columnwidth]{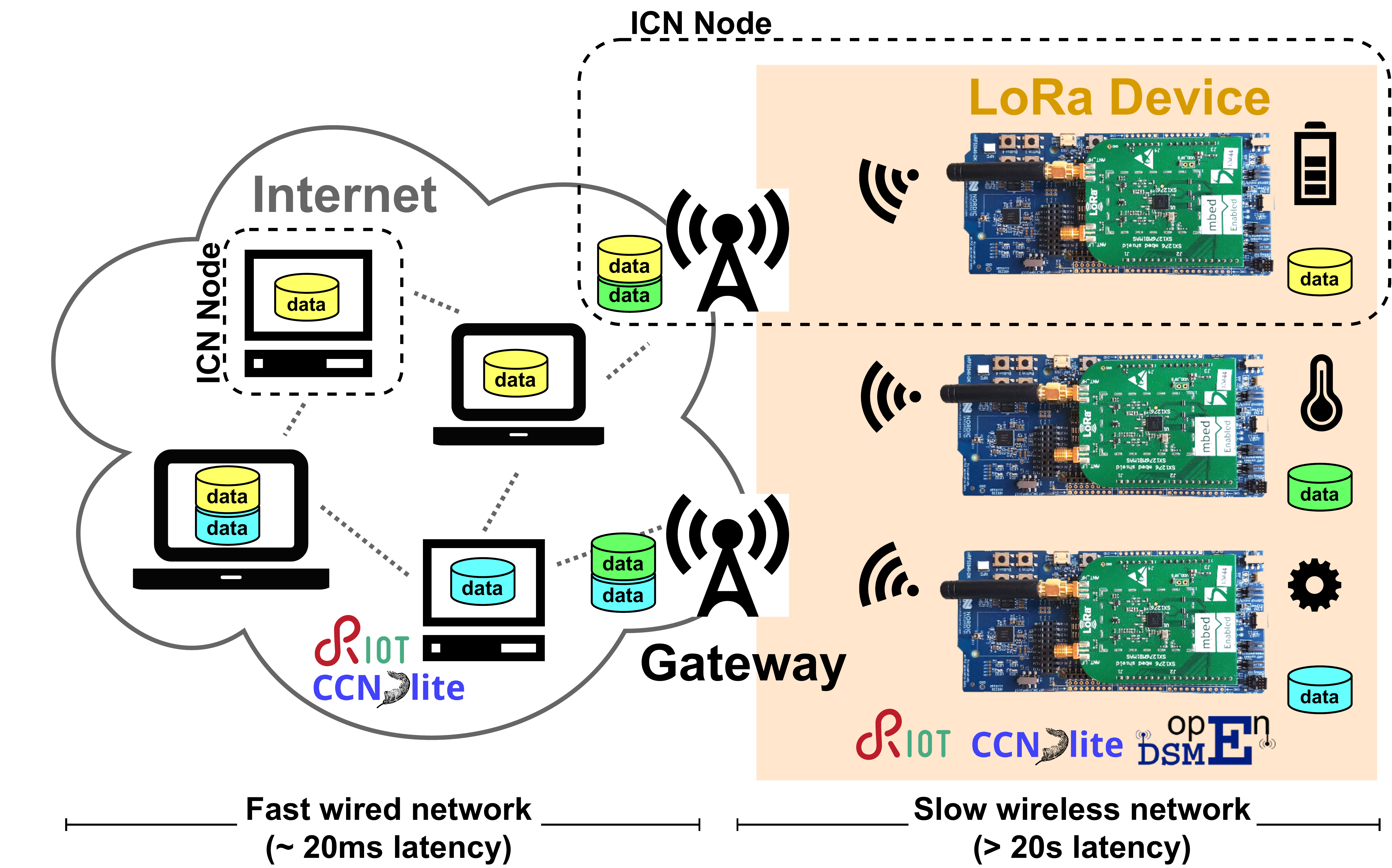}
  \caption{LoRa-ICN network and time domains.}
  \label{fig:motivation}
\end{figure}

LoRaWAN~\cite{lorawan-spec-11} provides a vertically integrated
network architecture for connecting LoRa networks and its constrained
devices to the Internet that is designed to offload power-constrained
wireless LoRa nodes as much possible:
gateways relay communication between the wireless link and network
servers (often co-located with additional application server
infrastructure) that manage the intricate energy-conservation regime
of connected LoRa devices.

The energy conservation objectives lead to a MAC layer design that
incurs dramatically higher latency and round trip times (RTTs) of
several seconds, compared to what connection-oriented Internet
transport protocols are typically designed to support. As a result,
LoRaWAN supports message-oriented transport through gateways and
dedicated network servers only, without a notion of end-to-end
communication from the Internet to LoRa nodes. While it is
theoretically possible to run bidirectional IP-based communication on
top of LoRaWAN~\cite{RFC-9011}, the resulting systems inherit
latency challenges of LoRaWAN for bidirectional communication that
would impact transport layer performance and applicability.

ICN has demonstrated benefits for improving data availability and
communication performance in constrained IoT
networks~\cite{bmhsw-icnie-14}. In this paper, we argue that ICN is
{\em also} a suitable network layer for connecting such challenged
edge networks to a more regular Internet, by leveraging hop-by-hop
transport functions,
ICN caching and minimal application-agnostic extensions.

Kietzmann~\etal~\cite{kaksw-liiel-22}
present a design of an improved, IEEE 802.15.4e DSME~\cite{IEEE-802.15.4-16} based MAC layer
for LoRa that supports packet-based communication, specifically
ICN-style Interest/Data communication.  Yet, RTTs can still be on the
order of seconds due to the underlying power saving regime.
Leveraging their work, we take an ICN-enabled LoRa subnet as a basis,
which is attached via an ICN forwarder on a gateway device. We develop
a delay-tolerant ICN communication framework that allows connecting
these LoRa sub-networks to a ``regular'' ICN Internet
(\autoref{fig:motivation}), with the following design goals: \one
supporting IoT sensor data transmission; %
\two supporting arbitrary orders of delays, without specific
assumptions of typical RTTs on other nodes on the ICN Internet;
\three not requiring application awareness on gateway nodes;
\four utilizing ICN-idiomatic communication to benefit from ICN principles such as accessing named data, Interest/Data semantics, caches, flow balance, etc.

We have developed interactions for IoT
communication use cases that leverage bespoke (but
application-agnostic) capabilities on gateway-based forwarders and the
{\em reflexive forwarding} extensions for ICN~\cite{draft-oran-icnrg-reflexive-forwarding}. These cases follow two patterns.
First, IoT sensor data retrieval from an Internet-based consumer using Interest/Data interactions; and
second, asynchronously ``pushing'' data from an IoT sensor to an Internet-based consumer with pub/sub semantics.

\noindent The contributions of this paper are the following:

\begin{enumerate}[wide, labelwidth=!, labelindent=0pt,itemsep=3.6pt]
  \item The design of delay-tolerant ICN-interactions and node behavior for this constrained environment.
  \item A complete implementation of the DSME MAC layer for LoRa
    ~\cite{kaksw-liiel-22} and our ICN protocol extensions on
    RIOT~\cite{bghkl-rosos-18}, serving common LoRa sensors and
    RIOT-based gateways~\url{https://github.com/inetrg/ACM-ICN-LoRa-ICN-2022.git}. %
  \item An experiment-based evaluation of the interactions on constrained IoT hardware, connected to an emulated ICN-Internet,
  and a comparison with vanilla ICN approaches.
\end{enumerate}

The rest of this paper is structured as follows:
\autoref{sec:background} describes essential LoRa and DSME background.
\autoref{sec:ps} discusses cor\-re\-spon\-ding challenges that our
system design, presented in \autoref{sec:sys_overview}, considers.
\autoref{sec:implementation} introduces our implementation, which
is the basis for an experimental evaluation in
\autoref{sec:evaluation}.
We discuss related work in \autoref{sec:related} and present our conclusions and future
work in \autoref{sec:conclusions}.

\section{Background}\label{sec:background}

In this section, we describe properties of the LoRa environment and the
DSME MAC layer that our work is based on.

\subsection{LoRa and LoRaWAN}\label{sec:background_lora}
LoRa defines a chirp spread spectrum modulation which enables a long
transmission range (kilometers), low energy consumption (millijoules)
at the cost of long on-air times. Duty cycle regulations further limit
the effective throughput (bits per second).  These features are still
attractive for many IoT use cases.  We operate on the EU 868\,MHz band
and configure a spreading factor 7, 125\,kHz bandwidth, code rate 4/5,
which results in a symbol time of 1.024\,ms.

LoRaWAN~\cite{lorawan-spec-11} is a popular system that operates on top of the
LoRa PHY. It defines a vertically integrated, and centralized network
architecture to integrate LoRa nodes to the IoT.  So-called network-
and application servers provide interfacing to the system.  A network
server interconnects applications and LoRa nodes, via gateways that
relay messages from and to the wireless network.  The network server
organizes MAC schedules centrally, while end devices operate in one of
three modes: class~A (intended for battery-powered devices) is purely
producer-driven, best-effort with very limited support for downlink
communication; class~C is not suitable for the low-power domain; and
class~B as a tradeoff between both.  LoRaWAN networks are subject to
collisions~\cite{f-cplal-17,ofjg-cccal-19} and scalability
issues~\cite{fbf-eslgc-18, ekbb-elcbe-20}. Class~B, albeit rarely
deployed, is designed to allow periodic downlink communication at low
energy, and exhibits reliability issues~\cite{mpp-edtpl-18,
  vht-idsl-19, rllcl-paodt-20}.  It further reveals long downlink
latencies. For example, Elbsir~\etal~\cite{ekbd-elced-20} measured an
average waiting time of 44\,s at 26\,\% delivery ratio in a relaxed
class~B configuration.  All those results motivate re-considering
the LoRa MAC system design.

\subsection{DSME and LoRa}\label{sec:background_dsmelora}
\begin{figure}
    \scalebox{.95}{\includegraphics{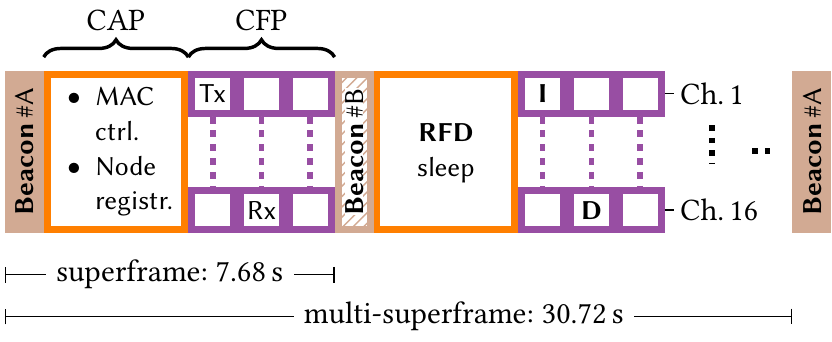}}
    \caption{Overview of the DSME multi-superframe structure. Perspective of a coordinator. Exemplary schedule for Interest (I) and data (D).}
    \label{fig:dsme}
\end{figure}

Motivated by the LoRaWAN deficiencies, we are basing our work on the new
DSME-based LoRa MAC design that was introduced by
\cite{kaksw-liiel-22}. It has the following key properties that are
relevant to this paper:

In DSME (\autoref{fig:dsme}), a coordinator emits beacons and initiates a synchronized
multi-superframe structure; beacon collisions are inherently resolved
for multiple coordinators in reach.  Constrained devices (RFD: reduced
function device) synchronize to that structure and join the subnet.
A superframe is separated into two periods for data transmission:
contention-access period (CAP), and contention-free period (CFP).
This time division facilitates battery powered nodes to enter
low-power mode periodically.  CFP slots assign unique and
frequency-multiplexed transmission resources between nodes to avoid
collisions, and provide a deterministic max. latency. Varying slot
assignments enable star-, peer-to-peer-, or clustered tree
networks. We focus on star topologies.

For the MAC we configure
\texttt{macSuperframeOrder}: 3, \\
\texttt{macMultisuperframeOrder}: 5, and
\texttt{macBeaconOrder}: 5.
This results in a slotframe structure of four superframes per multi-superframe,
a beacon interval and multi-superframe duration of 30.72\,s (applying the LoRa symbol time of 1.024\,ms from~\autoref{sec:background_lora}),
and provides 28 time slots $\cdot$ 16 frequency channels = 448 exclusive transmission cells.
Other slotframe structures trade off subnet size, throughput, energy, and latency; the latter can increase to over 122\,s in certain configurations~\cite{aksw-dslrc-22}.

\section{Problem Statement}\label{sec:ps}

DSME enables an improved LoRa MAC layer design for reliable
bidirectional communication, and it can be configured to provide
lower latencies compared to LoRaWAN. As such, it is a much better
basis for any packet-based higher-layer network stack, including ICN.
Still, due to the energy-conservation objectives and the properties of
the underlying LoRa PHY layer, even DSME incurs significant
delays for interactive communication, based on its multi-superframe
structure. These latencies (30 seconds or more) impose
significant challenges to any ICN Interest/Data communication, for
example, fetching a sensor value from a LoRa sensor, and will
require a delay-tolerant communication system.

Superficially, it seems straight-forward to add delay tolerance to
ICN, \eg by simply adding a face implementation for the DTN (Delay Tolerant Networking) bundle
protocol~\cite{RFC-5050} or by implementing a delay-aware forwarding strategy on a
forwarder. In reality, NDN~\cite{zabjc-ndn-14}- and CCNx~\cite{RFC-8569}-style ICN provides challenges
for inter-connecting networks with vastly different RTTs, which is
mostly due to the dual functions that Interests provide:

\begin{enumerate}[wide, labelwidth=!, labelindent=0pt,itemsep=3.6pt]
  \item Interests and Interest sending rates are central in the
    transport layer control loop of ICN receiver-driven transport
    services, \ie the Interest rate controls the throughput.
    Interests are used to trigger data transmissions in the first
    place, and to trigger retransmissions in case no corresponding
    Data messages have been received within a certain time interval.
  \item Pending Interests are temporary state in forwarders that is
    needed to implement a symmetric forwarding property in ICN, \ie to
    record the downstream face that corresponding Data messages should
    be forwarded on. A secondary function of pending Interest state it
    to enable {\em Interest aggregation} -- a feature that would
    prevent multiple Interests for the same Data object to be forwarded
    on the same path (when there is current pending Interest for that
    Data object). {\em Interest aggregation} effectively means {\em Interest
      suppression} for all but the first Interest that has been
    received by a forwarder in a certain epoch -- the Interest
    lifetime in the Pending Interest Table (PIT) of that forwarder.
\end{enumerate}

For achieving a reliable and decently performing communication
service, Interest state on forwarders {\em has to} expire, otherwise
Interest retransmission would {\em always} be suppressed by on-path
forwarders that have pending Interest state (and have not received the
corresponding Data object yet). There is a time relationship between
the Interest lifetime on forwarders and consumer retransmission
timers. For good performance, the Interest lifetime needs to be
shorter than the retransmission timer.

To cater to delay-prone networks, one could increase both values,
maintaining this property. In a heterogeneous network
environment (like the Internet), however, it is impossible to decide on ``good
values''. When connecting a high-RTT edge network to a high-speed and
low-RTT Internet, both the Interest lifetime and the Interest
retransmission timer would need to be adjusted for the end-to-end path
RTT.
Alternatively, adaptive suppression mechanisms in forwarders (\eg implemented in NFD~\cite{aszzm-ndg-21}) allow for Interest retransmissions in the presence of matching PIT entries. This does, however, still not solve the problem of guessing suitable timeout values for long and vastly different RTT and adopting these timers on every forwarder.
Future research and experiments should further investigate different options.

NDN Interests can provide an optional {\tt InterestLifetime} field
that allows a consumer to request more suitable Interest lifetime
durations (other than the 4 seconds default). We argue that this is
not likely to work well in actual deployments:

\begin{enumerate}[wide, labelwidth=!, labelindent=0pt,itemsep=3.6pt]
  \item Non-edge, high-speed forwarders are not likely to honor
    non-standard {\tt InterestLifetime} values for individual
    Interests to avoid the per-packet performance penalty.
  \item In DTN scenarios, RTTs and thus consumer-defined \\
  {\tt InterestLifetime} values could be significantly higher than 4
    seconds, and a core router may just object to spend memory
    resources for storing many Interests for a longer time.
  \item In DTN scenarios, the RTT may also change unpredictably,
    depending on caching, opportunistic contacts, new routing state
    \etc so the {\tt InterestLifetime} and the consumer Interest
    expiration time would have to be adapted constantly, which could
    introduce brittleness and inefficiency.
\end{enumerate}

It should be noted that ICN in-network congestion control and specific
per-forwarder strategies (for example, delay-tolerant forwarding
strategies) do not fundamentally resolve these issues because of
the interaction with consumers in the non-challenged network and their
different understanding of RTTs and retransmission timers.
We argue that, instead of guessing suitable {\tt InterestLifetime}
values and hoping for all on-path forwarders to honor the
corresponding Interest field, it is better to deal with varying and dramatically
higher RTTs (\eg in DTN scenarios) explicitly, with bespoke ICN
protocol mechanisms, without interfering with the ICN network layer
Interest lifetime.

\section{System Overview}\label{sec:sys_overview}

\begin{figure}
    \centering
    \subfloat[Delay-tolerant Data Retrieval like RICE.]{\scalebox{.9}{\includegraphics{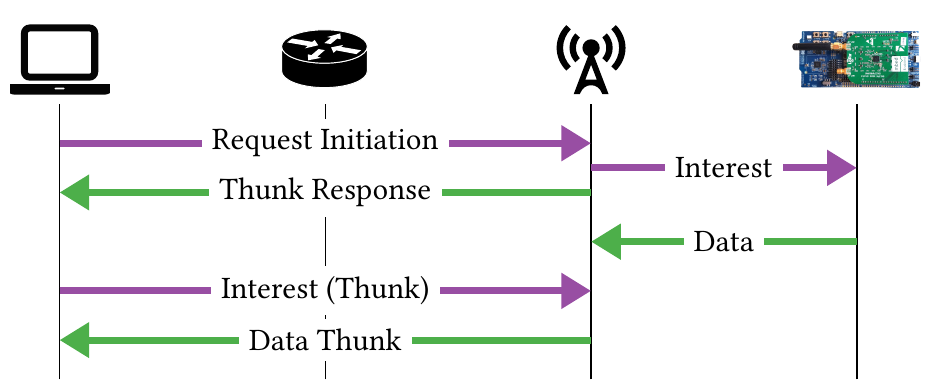}\label{fig:rice}}}\hfill
    \subfloat[Reflexive Push like {\em phoning home}.]{\scalebox{.9}{\includegraphics{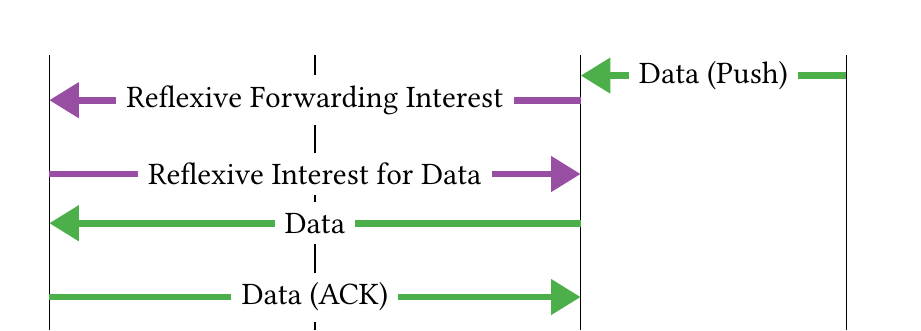}\label{fig:phone-home}}}\hfill
    \caption{Delay-Tolerant ICN.}
    \label{fig:abc}
\end{figure}

\autoref{fig:motivation} illustrates our system model: we want to
provide ICN delay-tolerant communication to edge networks, such as a
LoRa networks so that hosts on the ``regular'' ICN Internet can
communicate (\eg request data) with hosts in the challenged LoRa
edge network, without requiring Internet hosts and forwarders to apply
special {\tt InterestLifetime} parameters and retransmission timers.

Our work is based on three components: \one a mapping of ICN to DSME,
\two gateway node requirements, and \three delay-tolerant ICN protocol
mechanisms for interconnecting challenged networks (including but not
limited to ICN/DSME/LoRa networks) to non-challenged networks --
aiming for a seamless integration from an application perspective.

\subsection{Mapping of ICN to DSME}\label{sec:sys_mapping}
DSME provides a contention-access period that is prone to collisions,
and a contention-free period (see~\autoref{sec:background}) requiring
a priori slot negotiation.  We exclude node association and dynamic
slot allocation from this work, as they are orthogonal to the
information-centric and delay-tolerant networking aspects. Evolving
\cite{kaksw-liiel-22}, we simplify the ICN-DSME mapping and use the
CAP only for node registration (see below), and the CFP for regular
network layer traffic since it guarantees exclusive media access.  For
the CFP traffic, we implement static scheduling. In bidirectional
communication, each Interest slot is followed by a data slot. Consequently, presuming data availability,
a request is answered withing the same superframe.
For unidirectional data push, a single slot is allocated per node.

\subsection{Gateway Node Requirements}\label{sec:sys_req}

In our system, a LoRa gateway is an application-agnostic, caching ICN
forwarder that connects the narrowband LoRa network to the Internet and follows ``regular'' ICN behavior (\ie routing) in the upstream direction.
Hence, upstream congestion is uncritical since we consider a broadband network as the default deployment.
Downstream congestion on the constrained last hop is handled by the buffering gateway.
In addition to regular ICN forwarding and caching, the gateway
leverages knowledge about expected delays on the LoRa network for
adjusting PIT expiry times and {\tt InterestLifetime} accordingly.
This PIT state naturally prevents Interest flooding on the wireless
medium, as long as it remains active. Caching, as in other ICN
scenarios, offloads (re-) transmissions of Interests and Data messages
from the wireless link and the constrained nodes. Moreover, the
gateway provides these two additional functions:

\paragraph{Node Registration}
LoRa nodes register at the gateway after association, \ie
synchronizing to and joining a network that is advertised by a
coordinator. Re-joining a possibly different gateway operates at the
order of one (or few) beacon intervals. Nevertheless, it allows for
mobile nodes.  An overloaded Interest packet by the node indicates its
prefix, which establishes a downlink FIB entry on the gateway
(see~\cite{acim-ndnia-14}), and the face contains MAC information how
to reach that node.  Nodes can only serve content under that
prefix. On success, the gateway confirms the registration with a data
ACK.  On a FIB face timeout, \ie registration expiry, DSME management
routines could assist indication (future work).

\paragraph{Local Unsolicited Data}
The gateway accepts unsolicited ICN Data messages from registered LoRa
nodes and acts as a custodian for these nodes. The corresponding
content objects are stored in its CS, and the gateway will respond to
corresponding Interest messages from the Internet.
Caching strategies manage content placement and timeouts for cache eviction. Although gateways are not constrained in memory, least recently used content items are overwritten in case of overflow.

\subsection{Delay-Tolerant ICN Protocols}
\label{sec:sys-proto}

\paragraph{Delay-tolerant Data Retrieval (Fig.~\ref{fig:rice})}
We want to provide end-to-end ICN communication from an Internet
consumer to a LoRa node, \ie to enable Internet hosts to request
arbitrary content objects or to trigger computation in a Remote Method
Invocation (RMI) scenario (future work). We leverage the concept of
RMI for ICN (RICE~\cite{khokp-rrmii-18}) that provides access to
static data and dynamic computation results, supporting vastly longer
data production/retrieval times. Upon receiving a RICE request
initiation Interest, the gateway initiates an Interest message to the
LoRa node, as depicted by Figure~\ref{fig:rice}. A so-called ``Thunk
Response'' contains an indication for the waiting time, leveraging
link-knowledge about the DSME configuration in the LoRa network.

\paragraph{Reflexive Push (Fig.~\ref{fig:phone-home})}
Data generation (\eg sensor sampling) in the IoT happens sporadically
and asynchronously in many cases, which challenges the
receiver-driven (``pull'') ICN-paradigm \cite{bgnt-ssndn-14}. The high LoRa
latency further motivates a producer-driven data flow in order to
avoid periodic polling. This is consistent with~\cite{kaksw-liiel-22}
who suggest a unidirectional data push for LoRa-ICN.  In this
scenario, nodes need to register (as described above) before being
authorized to push content to the gateway, using the {\em Local
 Unsolicited Data} method.
This approach assumes a provisioned name as the {\em phoning home} destination that could be configured when registering the node at the gateway.

We forward these messages to a node on the Internet by leveraging the
{\em phoning home} use case of the {\em reflexive forwarding} extension to
ICN~\cite{draft-oran-icnrg-reflexive-forwarding}: the gateway sends an
Interest to a configured node on the Internet, which triggers a {\em
  reflexive Interest} by that node to retrieve the content object (Figure~\ref{fig:phone-home}).

This approach halves the number of resource-intensive wireless transmissions on the last hop,
and doubles the number of available DSME slots per multi-superframe.
It should be noted that a next-hop signaling does not introduce new security threats, since a network layer can never prevent a malicious neighbor from transmitting unwanted messages (or jamming) on the local link.
The slot-based MAC, however, naturally assists prevention of DDoS, triggered by
publishing LoRa nodes. A malicious node can simply be muted by the
coordinator (\ie the gateway), de-allocating its CFP slot.

\noindent\textbf{Note:}
We focus on communication aspects of the protocol mechanisms.
Security and corresponding configuration are out of scope for this paper.
Hence, we have slightly
simplified the protocol operations in our implementation of these
schemes (\autoref{sec:impl-proto}), \eg we do not use the RICE
request parameter retrieval for {\em Delay-tolerant Data Retrieval}.

\newcommand{\base}{{Vanilla (1)}\xspace}%
\newcommand{\smartc}{{Vanilla (2)}\xspace}%
\newcommand{\eagerf}{{Vanilla (3)}\xspace}%
\newcommand{\gwcust}{{Delay-tolerant retrieval}\xspace}%
\newcommand{\push}{{Reflexive push}\xspace}%

\section{Implementation and Deployment}\label{sec:implementation}

We describe our system implementation in \autoref{sec:impl-sys} and
the protocol implementation in \autoref{sec:impl-proto}.

\subsection{System Setup}
\label{sec:impl-sys}

\begin{figure}
    \includegraphics{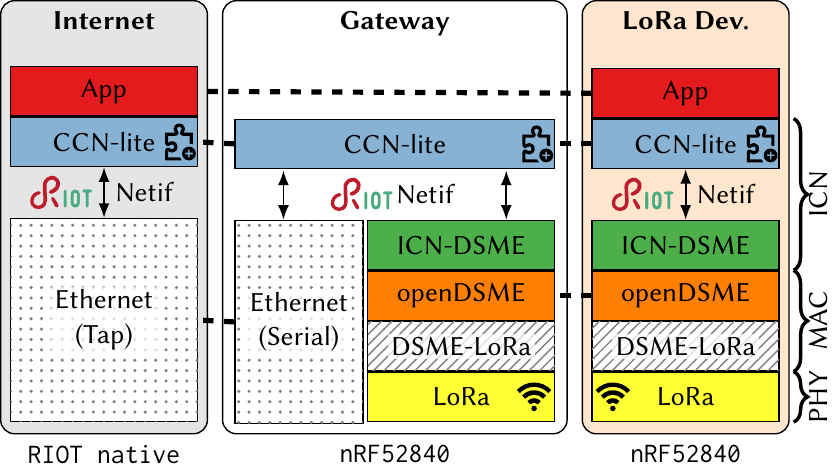}\hfill
    \label{fig:stacks}
    \caption{LoRa-ICN stacks on different devices with varying resources and network latencies.}
    \label{fig:impl_setup}
  \end{figure}

We have implemented this system on actual common off-the-shelf LoRa nodes, and we built our LoRa-ICN gateways on the same constrained hardware, to reduce implementation overhead. In a real-world LoRa network (\cf LoRaWAN), however, these gateways are not constrained in energy, memory, or processing
power and can serve many low-end nodes simultaneously, through radio concentrators.
LoRa devices, gateways, and Internet nodes operate the same ICN stack, to overcome incompatibility issues.
In the following, we describe the framework (\autoref{fig:impl_setup}) that we have created for experimentation.

\paragraph{RIOT~\cite{bghkl-rosos-18}}
We base our implementation on RIOT~2022.04. The networking subsystem
(namely GNRC) integrates CCN-lite as an ICN stack, which utilizes the
generic network interface layer (RIOT Netif
in~\autoref{fig:impl_setup}) to send and receive packets.  Currently,
wired Ethernet and 802.15.4 CSMA/CA wireless interfaces are available.
RIOT supports > 230 IoT boards and a \textit{native} port to execute
in a Linux process; it utilizes virtual TUN/TAP interfaces for
communication.
To build CCN-lite based gateways in RIOT that provide both, a fast
wired link and a slow long-range radio, we extend the OS integration
layer to utilize multiple network interfaces of varying types, behind
an ICN face.

\paragraph{CCN-lite~\cite{ccn-lite}}
Our integration bases on the latest version, checked out by RIOT 2022.04. CCN-lite provides an ICN forwarder implementation and common data structures: PIT, FIB, and CS. A hop-wise retransmission mechanism re-sends a pending Interest after a pre-configured timeout. Note, received Interest retransmissions will be aggregated when hitting an active PIT entry. PIT state expires after a pre-configured {\tt InterestLifetime} value, as usual.
We extend CCN-lite by runtime configuration abilities to adjust the PIT- and retransmission timeout, and the number of retransmissions dynamically.
Furthermore, we extend the core forwarder by protocol extensions (\includegraphics[width=.35cm]{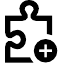}) described in~\autoref{sec:sys_overview} and the mapping to DSME (ICN-DSME in~\autoref{fig:impl_setup}).

\paragraph{openDSME~\cite{kkt-rwmnd-18}}
The open access DSME implementation for 802.15.4 radios was ported to
RIOT by Alamos~\etal~\cite{aksw-edmls-22} who also developed an
adaptation layer for LoRa (DSME-LoRa
in~\autoref{fig:impl_setup}). Their code is publicly available, albeit
not on RIOT upstream. We base our work on their implementation and add
interfaces to dynamically control MAC parameters (\ie ACK request,
send period) on a per-packet basis, through the RIOT network
interface. The southbound interface utilizes the 802.15.4 radio
abstraction API of RIOT.

\paragraph{LoRa Device}
We deploy the long-range sensor application on common low-power IoT hardware. The Nordic nRF52840 development kit consists of an ARM Cortex-M4 which provides 256\,kB RAM, 1\,MB flash, and runs at 64\,MHz. A SX~1276 LoRa radio shield is attached via pin headers and connects the external radio via SPI.
An adjusted transceiver driver implementation exposes the device an 802.15.4 radio, with LoRa specific timing parameters. This facilitates its usage with openDSME.
The sensor node is operated as a reduced function device and synchronizes to the DSME multi-superframe, indicated by a coordinator.
Afterwards, the node registers its ICN prefix using Interest/Data (see~\autoref{sec:sys_req}).

\paragraph{Gateway}
To reduce implementation overhead, we deploy our gateway on the same
hardware as the sensor application.
Our gateway acts as a
coordinator for LoRa nodes and creates the DSME slotframe structure through the wireless interface.
To communicate with a `fast' infrastructure ICN network in parallel (see forwarder and consumer below), we enable a second network interface; \texttt{ethos} is a RIOT specific
implementation for Ethernet over serial communication lines.
This is required because our experimentation platform lacks Ethernet hardware.
Real-world gateways, however, would simply use a gigabit Ethernet link.
Our serial device connects to a common Linux based workstation which
bridges to a virtual TAP bridge.

\paragraph{Internet (Forwarder and Consumer)}
Nodes on the Internet are emulated by RIOT-{\em native} instances to utilize
the same ICN stack, and connect to the same virtual TAP bridge as our
gateway. We deploy two nodes in a line topology, one forwarder and one
consumer. Both run in a \textit{Mininet}~\cite{mininet-framework-22} emulation to enable short
link delays of 20\,ms and optional link losses on the virtual wire.

\subsection{Protocols for Data Retrieval}
\label{sec:impl-proto}
\begin{figure*}
    \centering
    \subfloat[Vanilla]{\scalebox{.8}{\includegraphics{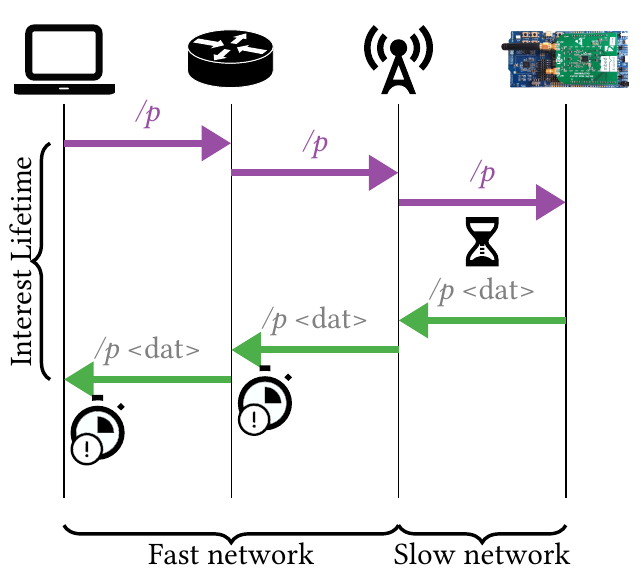}\label{fig:sequence_vanilla}}}
    \subfloat[\gwcust]{\scalebox{.8}{\includegraphics{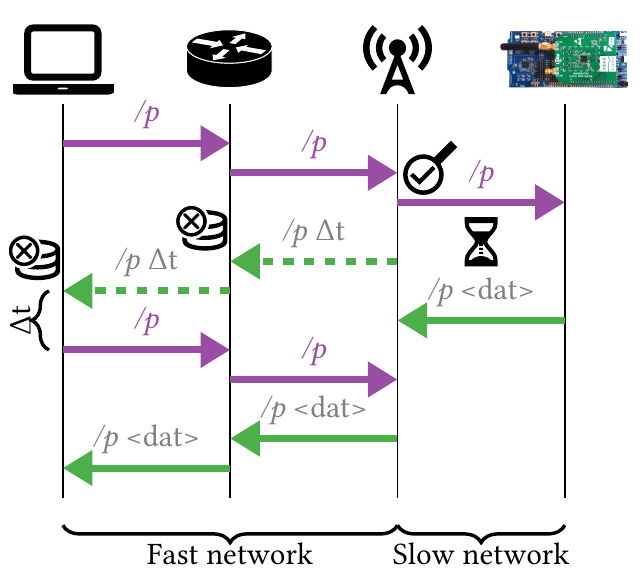}\label{fig:sequence_gw}}}
    \subfloat[\push]{\scalebox{.8}{\includegraphics{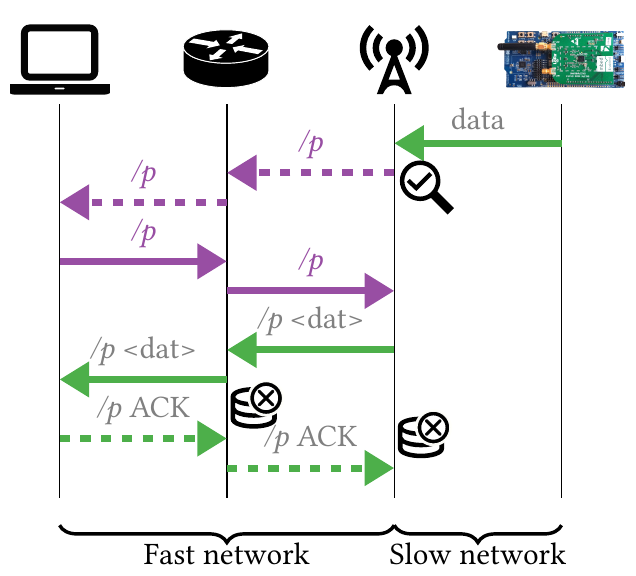}\label{fig:sequence_pushrfx}}}\hfill
        \scalebox{.8}{\includegraphics{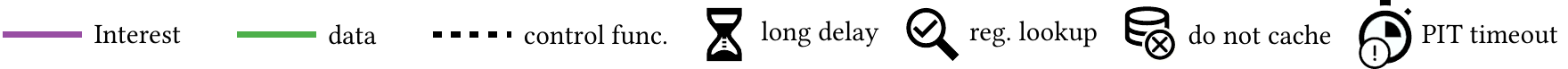}\label{fig:sequence_legend}}
    \caption{Sequence flows of Interest/Data and ICN extensions between nodes of different time domains. Data flows from LoRa producer to Internet consumer, either initiated by the consumer (\ref{fig:sequence_vanilla}-\ref{fig:sequence_gw}) or by the producer (\ref{fig:sequence_pushrfx}).}
    \label{fig:sequences}
\end{figure*}

We evaluated our system design, comparing its performance to that of
regular ICN Interest/Data communication.
To that end, we have defined three different data retrieval classes
corresponding to \autoref{sec:sys-proto}:

\begin{itemize}
\item {\em Vanilla ICN Request} for regular Interest/Data interactions
    initiated from a consumer on the Internet;

\item {\em Delay-tolerant Data Retrieval} using a
      simplified RICE exchange initiated from a consumer on the Internet;

\item {\em Reflexive Push} using {\em reflexive forwarding} and the {\em phoning home}
  use case initiated from the producer.
\end{itemize}

\paragraph{Vanilla ICN Request}
We assume a regular Interest request from the Internet to the LoRa
sensor (Figure~\ref{fig:sequence_vanilla}).
The request faces a non-typical long round trip time at the gateway,
conflicting with PIT state on forwarders.  \one ``Regular'' forwarders
that are not aware of the long delay domain are likely to operate on a
fast-network timescale.  PIT state that expires before data arrival
prevents forwarding on the reverse path.  \two Interest
retransmissions are common in ICN, albeit left to transport or
application layer implementations.
In general, regular ICN-based data retrieval quickly leads to polling
and unterminated retransmissions when facing long delays.  Two
built-in ICN countermeasures are worth discussing: first, {\tt
  InterestLifetime} dictates the PIT entry expiration time on
forwarders. Increasing {\tt InterestLifetime} solves the problem of
expired PITs, however, it also requires forwarders to maintain state
during the long DSME-LoRa round trip. In addition to occupying PIT
memory, this approach affects Interest retransmissions (as a response
to timeouts at consumers). Second, common ICN implementations (\eg
RICE, NFD~\cite{aszzm-ndg-21}) rely on consumer-based retransmissions
(contrasting in-network retransmissions). This {\em requires} PITs
to expire fast, otherwise, a retransmission will be suppressed.

\paragraph{Delay-tolerant Data Retrieval}
We have implemented the interaction from
\autoref{sec:sys-proto} by adding server logic to the link-aware gateway that
is triggered by the reception of corresponding Interest messages from
consumers in the non-challenged Internet (Figure~\autoref{fig:sequence_gw}).
The gateway performs three major actions after an incoming Interest:
\one It first checks for a registered LoRa node that falls under the
requested prefix, in its FIB.  \two On a missing FIB entry, it
immediately returns a data NACK.  \three On success, it forwards the
Interest as per regular forwarding using the FIB face towards the LoRa
node.  On forwarding, the gateway replies with a distinct data NACK
(we call it WAIT) which contains an estimated data arrival time. A gateway can provide accurate estimates in the future, using its knowledge of the DSME configuration upfront, the internal scheduler state, as well as the current traffic load (queue length).
This data packet
satisfies the initial Interest, corresponding in-network state, and terminates
potentially inappropriate ICN-based retransmissions.  The estimated
data arrival time enables the consumer application to set an
appropriate retry timer, without the need for specific producer knowledge and varying long delays introduced by DSME-LoRa.
NACK/WAIT data packets in \two and \three
must not be cached, though, to prevent serving a subsequent request of
the same name from the CS.  Finally, after a repeated Interest
request, the data item is likely served from the gateway.

\paragraph{Reflexive Push}
Our protocol flow (Figure~\autoref{fig:sequence_pushrfx}) implements
the second interaction from~\autoref{sec:sys-proto}.
It suggests two
nested Interest/Data exchanges.  After successful content placement on
the gateway, using {\em Local Unsolicited Data}, this one indicates data by sending an Interest packet
that contains the data name, to the consumer. An additional packet
indicator triggers the establishment of a temporary downlink FIB entry
on forwarders for that specific name, which points to the incoming
face.  The consumer can return a {\em reflexive Interest}, requesting the
announced data; it follows the previously established FIB path. Data
is served from the gateway cache as usual, satisfying PIT state on the
reverse path, and additionally removes the temporary FIB entries.  An
optional final data ACK terminates the initial Interest request.

\section{Evaluation}\label{sec:evaluation}

We describe experiment configurations in \autoref{sec:eval-config},
measurement results for protocol performance in
\autoref{sec:completion},  results from our analysis of communication overhead in
\autoref{sec:overhead}, and system overhead of the protocol stacks in \autoref{sec:impl-overhead}.

\subsection{Experiment Configuration}
\label{sec:eval-config}

We conducted five experiments (comparing our two schemes described in
~\autoref{sec:sys_overview} with three {\em Vanilla ICN} variants):

\begin{description}
  \item[\base] Baseline scenario with unchanged ICN and common parameter settings.

  \item[\smartc] Delay-aware consumer with extended\\{\tt InterestLifetime} and retransmission interval.

  \item[\eagerf] Like \textbf{(2)}, additionally forwarders observe the long {\tt InterestLifetime} and set their PIT timer accordingly.

  \item[\gwcust] Gateway acts as a special proxy for long-delay producers and returns a distinct re-try instruction on first request.

  \item[\push] Producer initiates a transaction by pushing data to
    gateway CS which triggers a reflexive Interest/Data interaction for
    retrieving content.
\end{description}

In our experiments, we use unique content names, prefixed with a LoRa
node ID and incremental local object counters.
Data contains either a
random integer value or an ACK, NACK, or WAIT instruction with a time
hint.
This fixed size scheme leads to a frame size of 31\,Bytes for Interest and 36\,Bytes for data, which leaves headroom to the maximum frame size of 127\,Bytes. Longer packets, however, could be compressed~\cite{gksw-innlp-19} and fragmented~\cite{lgsw-cdsfr-20} in the future.
Every content item is requested/indicated once, with an average
interval of one minute (60$\pm$10\,s uniformly distributed).  For a
fair comparison between consumer- and node initiated traffic, we
produce sensor data on the LoRa node after an incoming Interest. Data
returns during the subsequent CFP slot within the same superframe
(compare~\autoref{sec:sys_mapping}). Our measurements include: \one
completion time, \ie the delay between issuing a transaction and data
arrival at the consumer; \two resilience, \ie the rate of successful
transactions; \three protocol overhead, \ie the number of transmitted
packets per content item.  Thereby, we deploy an idealized scenario
with 0\% -- and the case for 5\% link loss on the Internet emulation.

\begin{figure*}
  \hfill{\includegraphics[width=.4\textwidth]{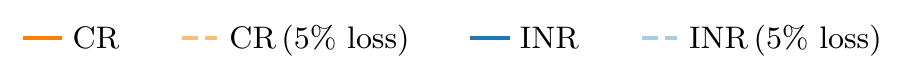}\label{fig:ttc_legend}}\hfill\vspace{-.5cm}
  \subfloat[\base]{\includegraphics[width=.2\textwidth]{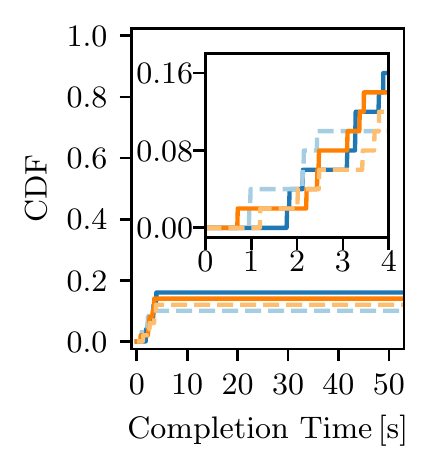}\label{fig:a1_a2}}
  \subfloat[\smartc]{\includegraphics[width=.2\textwidth]{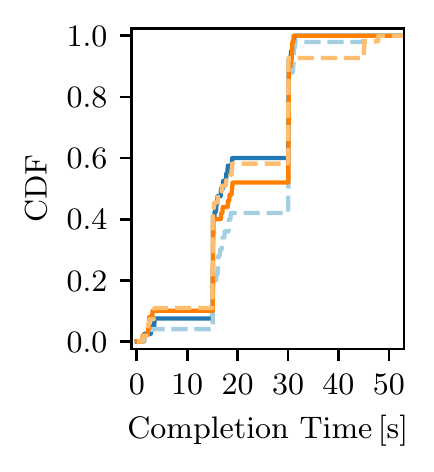}\label{fig:b1_b2}}
  \subfloat[\eagerf]{\includegraphics[width=.2\textwidth]{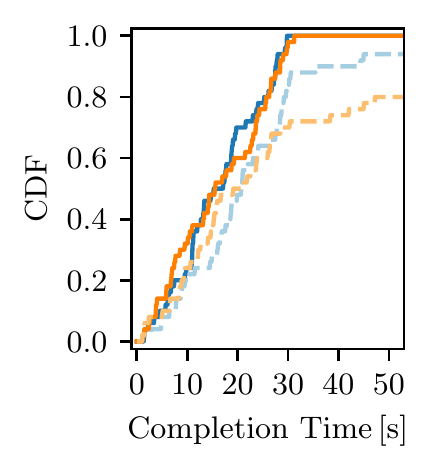}\label{fig:c1_c2}}
  \subfloat[\gwcust]{\includegraphics[width=.2\textwidth]{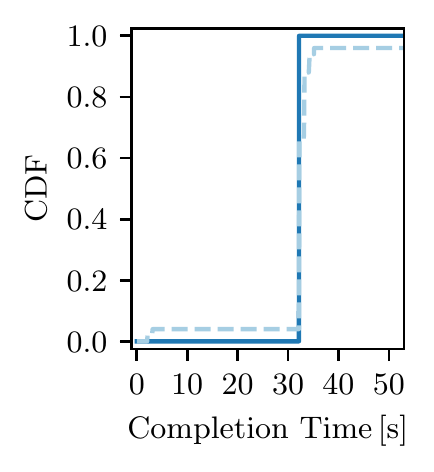}\label{fig:gw}}
  \subfloat[\push]{\includegraphics[width=.2\textwidth]{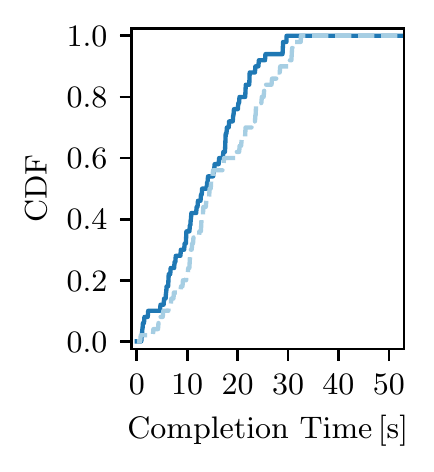}\label{fig:push}}\hfill
  \caption{Time to content arrival with long producer delays. Vanilla ICN in varying configurations and our extensions employ in-network retransmissions (INR) or consumer retransmissions (CR), and we vary the link loss.}
  \label{fig:ttcs}
\end{figure*}

\autoref{tbl:scenario_overview} summarizes our parameter settings.
All but the last scenario require the gateway and node to lift the PIT
expiration time to the long delay domain. We conservatively chose
60\,s which reflects $\approx$ two times the multi-superframe duration
of the MAC (compare~\autoref{sec:background_dsmelora}). Retransmits on the LoRa
hop are disabled since we utilize exclusive CFP resources.

For Vanilla ICN, we distinguish the case with in-network
retransmissions (INR) and consumer-based retransmissions (CR), with
different PIT timeout behavior.  Our \textbf{\base} configuration
assumes that Internet nods are unaware of the long delay
domain. Hence, we set a PIT expiration time of 4\,s according to
default settings of the common NFD implementation~\cite{aszzm-ndg-21}
and enable three network layer retransmits, each after 1\,s, which
reflects the initial round-trip estimation of TCP~\cite{RFC-6298}.  In
\textbf{\smartc}, a consumer is aware of long producer delays, hence,
we set {\tt InterestLifetime} and PIT expiration time to 60\,s as
well, and adjust the retransmission interval to 15\,s. Forwarders do
not adopt the long timeout value.  In \textbf{\eagerf} the forwarder
adopts the {\tt InterestLifetime} value of the incoming packet and
sets its PIT expiration time accordingly, \ie to 60\,s. This does not
change its retransmission behavior in the INR case, however.  We
present two alternative solutions: \textbf{\gwcust} gets along with
`short' \base parameters and utilizes INR.  \textbf{\push} inverts the
original ICN semantic and consists of two nested Interest/Data flows
that utilize `short' time parameters analogously.

\begin{table}[]
  \centering
  \caption{Scenario and parameter overview including four measured nodes. (Abbreviations: INR=In-network retransmission, CR= Consumer retransmission, \xmark= not applicable).}
  \label{tbl:scenario_overview}
  \begin{adjustbox}{max width=1\columnwidth}
    \setlength{\tabcolsep}{3pt}
    \begin{tabular}{l  r  r  l  r l r l r c }
        \toprule
        \multicolumn{2}{c}{\multirow{2}{*}{\textbf{Scenario}}}
        &\multicolumn{2}{c}{\textbf{Cons.}}
        &\multicolumn{2}{c}{\textbf{Fwd.}}
        &\multicolumn{2}{c}{\textbf{Gw.}}
        &\multicolumn{2}{c}{\textbf{Node}}
        \\
        \cmidrule(lr){3-4}
        \cmidrule(lr){5-6}
        \cmidrule(lr){7-8}
        \cmidrule(lr){9-10}
        &
        &\includegraphics[height=.425cm]{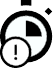}&\includegraphics[height=.375cm]{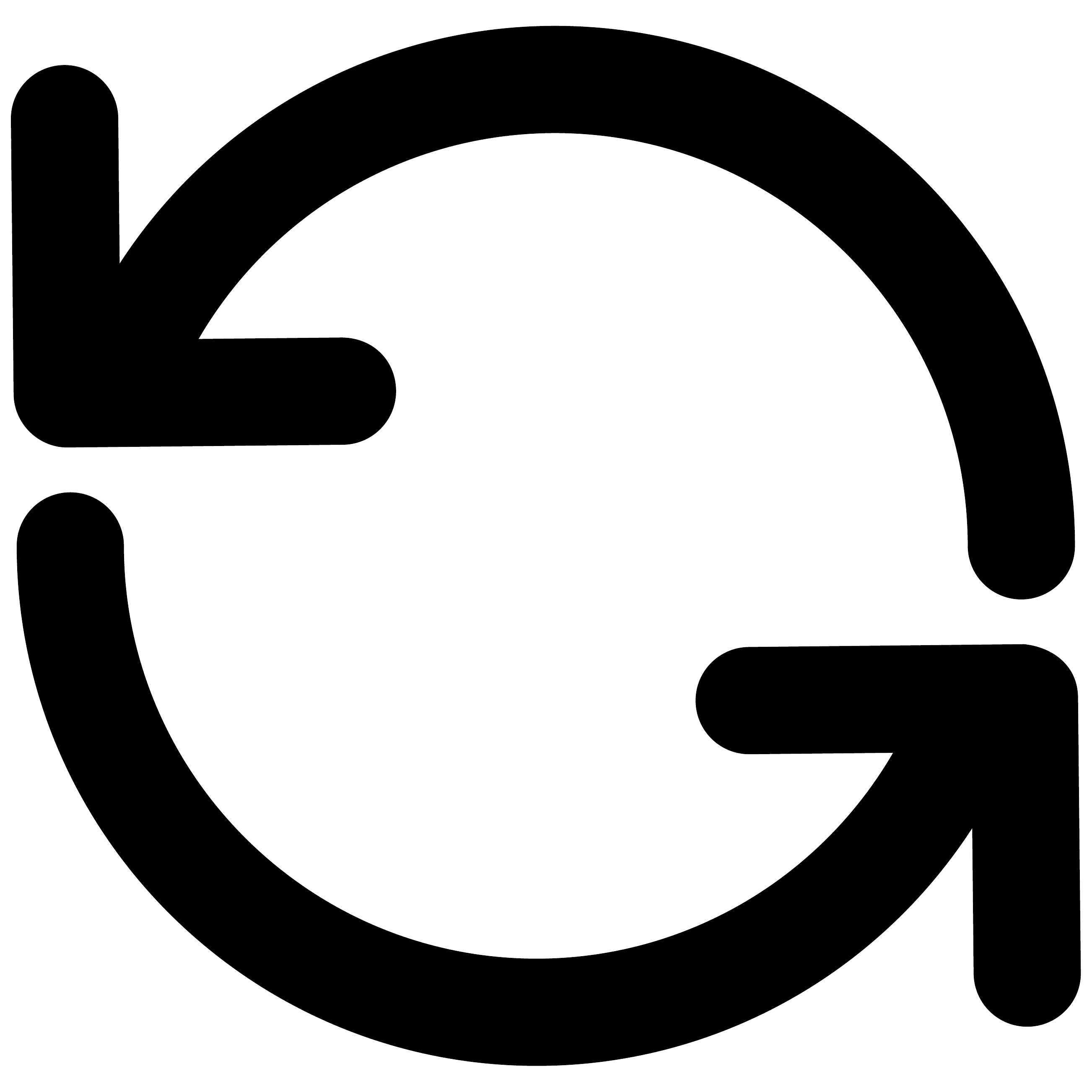}
        &\includegraphics[height=.425cm]{figures/timeout}&\includegraphics[height=.375cm]{figures/retrans}
        &\includegraphics[height=.425cm]{figures/timeout}&\includegraphics[height=.375cm]{figures/retrans}
        &\includegraphics[height=.425cm]{figures/timeout}&\includegraphics[height=.375cm]{figures/retrans}
        \\
        \midrule
        \multirow{2}{*}{\textbf{\base}}&
        INR&4&3:1&4&3:1&60&0:0&60&\xmark
        \\
        \cmidrule(lr){2-10}
        &
        CR&4&3:1&4&\makecell[c]{\xmark}&60&0:0&60&\xmark
        \\
        \midrule
        \multirow{2}{*}{\textbf{\smartc}}&
        INR&60&3:15&4&3:1&60&0:0&60&\xmark
        \\
        \cmidrule(lr){2-10}
        &
        CR&60&3:15&4&\makecell[c]{\xmark}&60&0:0&60&\xmark
        \\
        \midrule
        \multirow{2}{*}{\textbf{\eagerf}}&
        INR&60&3:15&60&3:1&60&0:0&60&\xmark
        \\
        \cmidrule(lr){2-10}
        &
        CR&60&3:15&60&\makecell[c]{\xmark}&60&0:0&60&\xmark
        \\
        \midrule
        \textbf{\makecell[l]{Delay-tolerant\\retrieval}}
        &
        INR\textsuperscript{\textbf{1}}&4&3:1&4&3:1&60&0:0&60&\xmark
        \\
        \midrule
        \textbf{\makecell[l]{Reflexive-push}}
        &
        INR&4&3:1&4&3:1&4&3:1&\makecell[c]{\xmark}&0:0
        \\
        \bottomrule
    \end{tabular}
    \smallskip
\end{adjustbox}
  \begin{minipage}{\textwidth}
    {\footnotesize \includegraphics[height=.425cm]{figures/timeout} PIT timeout [s] \quad \includegraphics[height=.375cm]{figures/retrans}  Retransmission attempts and timeout [\#:s]}\\
    \vspace{.2cm}
    {\footnotesize\textsuperscript{\textbf{1}}Additional retry based on WAIT instruction on first request.}
  \end{minipage}
\end{table}

\subsection{Completion Time and Resilience}\label{sec:completion}
\autoref{fig:ttcs} presents the cumulative distributions of completion times of successful transactions.
These values are mainly affected by the multi-superframe duration of the MAC (30.72\,s) which dictates the maximum latency of a unidirectional long-range transmission between.
Data losses result in infinite completion times, hence, the end value of each graph inherently reflects its success ratio.

\paragraph{\base (Fig~\ref{fig:a1_a2})}
10--16\%\, of requests are successful and finish in less than the PIT
timeout of 4\,s.  This is the case for Interest that happen to arrive
at the gateway short before a DSME transmission slot occurs.  Link
losses further drop the success rate by 2--6\%, but different
retransmission pattern do not provide a significant effect.

\paragraph{\smartc (Fig.~\ref{fig:b1_b2})}
Completed transmissions in <4\,s resemble properties of the \base case.
Steps at 15\,s indicate the poll interval of the consumer, which recovers losses from long DSME-LoRa delays. This requires, however, that forwarder PIT state
expires fast (here 4\,s) to prevent {\em Interest aggregation}.
Losses delay the completion and are compensated faster with INR overall ($\approx$ 32\,s), though, CR recovers 20\,\%  more requests on the first retry. Conversely, 10\,\% of the requests require a third retry with CR, to complete successfully ($\approx$ 45\,s).
A comparison of CR with and without loss reveals a diverse picture. Here, the lossless case surprisingly satisfies fewer requests after the first retry, which is an effect of randomized experimental requests.

\paragraph{\eagerf (Fig.~\ref{fig:c1_c2})}
Cases without link loss require 32\,s at max. (multi-superframe duration) to retrieve all content, which directly reflects the delay distribution of the DSME-LoRa MAC.
The long PIT state on both consumer and forwarder allow data forwarding whenever it is ready, reflecting the case for soft-state subscription by a long-lived Interest~\cite{cpw-cpsni-11}.
Link losses, however, demonstrate the drawback of this approach.
CR prevent effective loss recovery while PIT sate is active on the forwarder, and drop the delivery rate to 80\,\%.
INR recover most losses and result in 94\,\% delivery.
This approach only performs well under the assumption that \one every forwarder adopts the long PIT timeout, and \two content can be retrieved within that time.

\paragraph{\gwcust (Fig.~\ref{fig:gw})}
Requests finish in almost exactly 32\,s in the lossless case, which is the returned WAIT time of the gateway after the first request of a content item.
This static worst case value could be reduced with a latency estimator model on the gateway, allowing for targeted completion times.
The gateway retrieves content from the node during WAIT, and caches it. A subsequent request of that item is answered from the gateway CS.
Additional link losses are mostly recovered by INR and perform similarly to the lossless case, however, two effects are noteworthy:
\one $\approx$ 5\,\% of the requests finish below 3\,s. A loss of the first data packet, which contains a WAIT instruction, triggers an INR which is already satisfied by the gateway.
\two $\approx$ 8\,\%  of the requests are not satisfied. Our implementation uses a short circular list of future requests to re-issue, which avoids (larger) PIT state over long time. In the loss case, entries stayed longer in the list and got overwritten occasionally, leading to un-requested data. In practice, the list should be provided with timeout values and dimensioned according to traffic load.

\begin{figure*}[]
    \centering
    \includegraphics[width=1\textwidth]{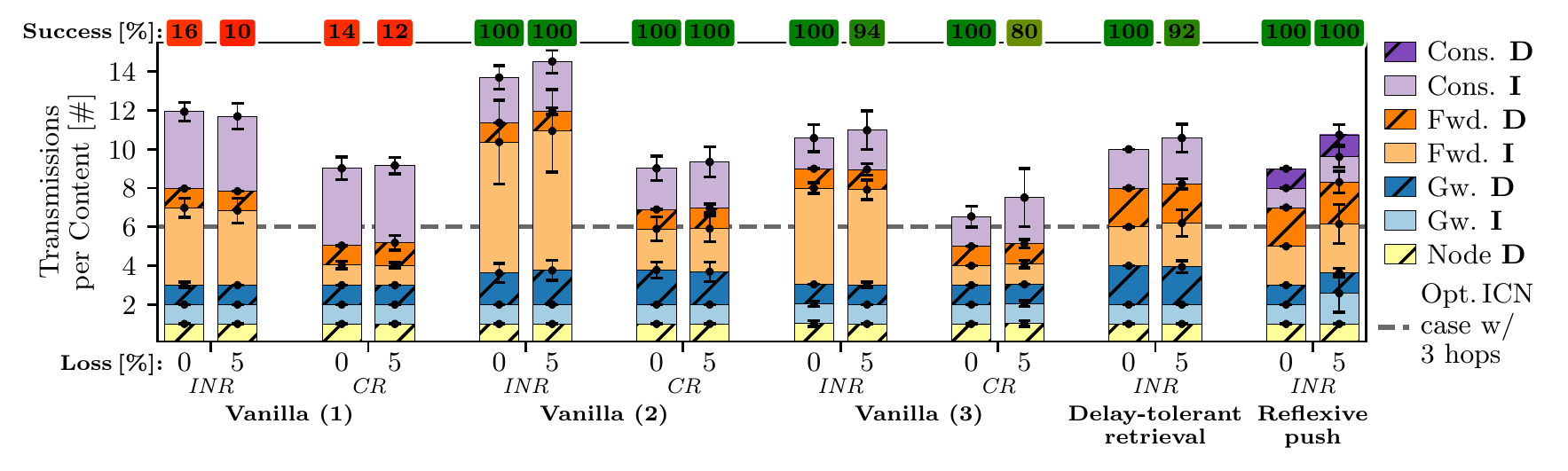}
	\caption{Transmissions per content item (protocol overhead) separated into consumer (Cons.), forwarder (Fwd.), gateway (Gw.), and LoRa node. Vanilla ICN in varying configurations and our extensions employ in-network retransmissions (INR) or consumer retransmissions (CR), and we vary the link loss. Lossless ICN transmission via 3 hops is indicated by the dashed line.}
    \label{fig:tx_per_content}
\end{figure*}

\paragraph{\push (Fig.~\ref{fig:push})}
Transactions finish in max. 32\,s (multi-superframe duration) with 100\,\% success. Completion times reflect the delay distribution of DSME-LoRa, similarly to the \eagerf scenarios. Herein, the additional round trip of a nested double Interest/Data flow has a negligible overhead when directed towards the fast network.
Thereby, losses are smoothly recovered by INR, at minimal time overhead.
Contrasting \eagerf, this approach works with arbitrary (producer) delays and forgoes the need to adopt long PIT timeout values on Internet nodes.

\paragraph{Findings}
Expired PIT state on the reverse path is the prevalent obstacle with
vanilla ICN and prevents round tips >4\,s, which renders the baseline
scenario unusable in this domain.
Application-aware consumers overcome long delays, however, the
performance heavily depends on the (arbitrary) choice of a poll
interval and is susceptible to {\em varying} delays.
Increasing the {\tt InterestLifetime} on the complete forwarding path,
instead, is challenging.  \one We cannot expect real forwarders to
blindly adopt arbitrary PIT timers.  \two Without in-network retransmission in
place, long-lived PIT state harms reliability.
The \gwcust case overcomes requirements of long PIT state and blind
polling. It thereby relieves Internet nodes and applications from
knowledge of the (variable) long time domain. Consumer implementations
become more complex, therefore.
A reversed transaction flow with \push facilitates efficient, reliable, and  `timely' transactions.

\subsection{Communication Overhead}\label{sec:overhead}

\autoref{fig:tx_per_content} quantifies the protocol overhead for
every node and scenario (cf. \autoref{sec:eval-config}) and shows the
number of transmitted Interest and Data packets per requested content
item as well as the success rate, replicated from~\autoref{fig:ttcs}.
In a three hop network, an optimal ICN request-response requires six
packets, as indicated by the dashed line.  Recall that all scenarios but {\em \push} lift the PIT
timeout on the gateway and disable network layer retransmissions on
the LoRa link, to preserve sparse resources.  Consequently, gateways
only transmit one Interest towards nodes that respond with one data
packet per request.

\paragraph{\base} These scenarios reveal notable overheads by futile retransmission, regardless of link loss. Up to two times as many packets are transmitted, compared to the ideal case, with little overall success.
With INR, both forwarder and consumer transmit at maximum (4 Interests/content), while CR keeps forwarder overhead low (1 Interest/content). Interests are aggregated as long PIT state persists. Standard retransmit intervals cannot cope with long delays.

\paragraph{\smartc} INR reveal the highest overhead among all scenarios (15 transmissions), sending requests at two timescales. Every consumer Interest is forwarded {\em and} retransmitted by the forwarder, regardless of the long delay of the producer.
In contrast, CR overhead ($\approx$ 9 transmissions) is on par with \base CR but satisfies all requests, without blind forwarding. Hence, PIT timeouts < consumer poll intervals that operate at the prevalent delay domain are a viable option for the conventional ICN paradigm. Short-lived PIT state cannot prevent duplicate data transmission by the gateway, though, when Data faces expired PIT state on a forwarder.

\paragraph{\eagerf} INR recovers link losses, while a `sufficiently' long PIT expiry time prevents consumer-based retransmissions. The CR case (without loss) thus operates with little overhead ($\approx$\,7.5 transmissions) but is not vital due to high sensitivity to link loss.

\paragraph{\gwcust} Our approach generally increases the required transmissions per content, introducing a second round trip between gateway and consumer. Hence, it performs optimal in the lossless case by transmitting 10 packets: 2xInterest/Data on fast nodes, and 1xInterest/Data on the LoRa link. INR marginally increase the overhead.
The total overhead compares to \smartc with CR, however, it surpasses
blind polling.

\paragraph{\push} Our second approach inverts the flow direction and introduces a second round trip between gateway and consumer as well. In contrast to \gwcust, however, only a single LoRa transmission is required to place producer content in the gateway cache. %
This producer oriented optimization results in an optimal number of 9 transmissions per content, reflected by the lossless case. INR increases up to 10.5 transmissions (avg) on loss.
Data ACKs by the consumer are optional and terminate the initial Interest of the gateway. Omitting these packets is principally possible to reduce transmissions, however, this conflicts with INR.

\paragraph{Findings}
\gwcust and \push are robust, operationally efficient, and can
tolerate varying delays of the DSME-LoRa MAC. In contrast, vanilla ICN
requests suffer from long and unpredictable delays. Naive consumer
polling is an inefficient but viable ICN-idiomatic alternative,
provided that Interests expire on the forwarding path and polling
intervals are set in agreement with practical delays.

\subsection{System Overhead}
\label{sec:impl-overhead}

We evaluate the resource overhead of our protocol stack and focus on the battery driven LoRa device, since gateways and Internet nodes are not resource constrained and remain unchallenged by common LoRa traffic.

\paragraph{Energy Consumption}
We present the energy consumption per multi-superframe in \autoref{tbl:scenario_overheads}, as well as the corresponding nodal lifetimes when operated from an off-the-shelf AA alkaline battery (2800\,mAh). Our results are based on extensive measurements performed in~\cite{aksw-dslrc-22}, which quantify the energy consumption for passive and active periods of the DSME-LoRa superframe structure.
Radio operations dominate consumption, \ie wireless transmission and (idle) reception.
To confirm this observation, we also measure the active  CPU time throughout our experiments, which is as low as $\approx$\,0.25\,\% for all protocols on the constrained node, and around 0.3\,\% on the gateway. The latter  increases with growing network sizes.

Vanilla ICN request values include the alternative operation without a MAC (ignoring wireless interference),  which strongly motivates the choice of a duty cycling MAC from the energy perspective. Without duty cycling,  the lifetime is limited to 10 days.
Enabling the MAC reduces the energy consumption by two orders of magnitude, which leads to a lifespan of 230\,days  in the vanilla ICN request and delay-tolerant data retrieval case, assuming that the gateway shields LoRa devices effectively from retransmits.
Reflexive push almost halves the energy consumption due to unidirectional transmission, which further increases the lifetime to more than a year.

\begin{table}[]
  \centering
	\caption{Energy consumption per multi-superframe and lifetime for the protocols under consideration.} 
  \label{tbl:scenario_overheads}
  \begin{adjustbox}{max width=1\columnwidth}
    \begin{tabular}{l  r  c }
        \toprule
        \textbf{Protocol} & \textbf{Energy [mJ]} & \textbf{Lifetime [d]}\\
        \midrule
        \textbf{Vanilla ICN request}  &  &  \\
        \quad{w/o MAC}  & 1247.46 & ~10\\
        \quad{w/ MAC}  & 51.42 & 230 \\
        \textbf{Delay-tolerant data retrieval} & 51.42 & 230\\
        \textbf{Reflexive push}  & 30.83 & 384\\
        \bottomrule
    \end{tabular}
    \smallskip
\end{adjustbox}
\end{table}

\paragraph{Memory Requirements}
Our network stack is runtime configurable to operate the three protocols for data retrieval (\autoref{sec:impl-proto}). Hence, the firmware image is the same for all configurations and requires 143\,kB in ROM (text + data segment) and in 19\,kB RAM (bss + data segment), almost half of which is occupied by openDSME.
The remaining RAM (256\,kB on nRF52840) is reserved for dynamic runtime memory allocation (heap). Both openDSME and CCN-lite utilize \texttt{malloc}, and we track the combined heap statistics which ranges between 6--8\,kB in all experiment runs. Thus, our LoRa-ICN stack can even be deployed on much smaller IoT hardware.

\section{Related Work}\label{sec:related}

\paragraph{Advancing LoRa(WAN)}
To overcome limitations of the centralized LoRaWAN architecture, multi-hop extensions for LoRa~\cite{tbs-eticl-17,gbv-slpld-18,bgbts-teelm-19,ck-lmnrc-20} have been proposed.
These are orthogonal to our work since we focus on single-hop topologies.

Contention-based~\cite{lbbp-calma-20, knwm-pecsl-20} and scheduled MAC layers~\cite{yte-talsa-19, zakp-ttlii-20, hoof-tlrri-20, kuf-ldlif-21} for LoRa indicate performance improvements compared to LoRaWAN.
Alamos~\etal~\cite{aksw-edmls-22,aksw-dslrc-22} re-utilize IEEE 802.15.4e DSME (Deterministic and Synchronous Multi-Channel Extension)~\cite{IEEE-802.15.4-16} to coordinate LoRa radios, with few modifications to the radio configuration.
Fixed time-slotted DSME paired with low data rates increases latencies even further, though.
In this work, we enable LoRa to run a robust DSME-based MAC layer with latencies that we are able to handle.

RFC~9011~\cite{RFC-9011} specifies Static Context Header Compression
and Fragmentation~(SCHC) for IPv6 over LoRaWAN.  We agree that
compression and fragmentation are crucial, but do not address the
latency issues for transport protocols.  Also, SCHC does not
fix the underlying MAC, which is prone to collisions and depends on
network server scheduling.

\paragraph{ICN and the IoT}
The IoT benefits from ICN~\cite{bmhsw-icnie-14,pf-britu-15,mwt-tucin-16,sblwy-ndnti-16,saz-dinps-16,gklp-ncmcm-18, aarl-raini-19, yzmmz-panbs-21}.
An important observation in prior work is that IoT~scenarios require the adaptation of the MAC layer to prevent unnecessary broadcast and preserve energy resources~\cite{kgshw-nnmam-17}.
Current analyses either base on 802.15.4 CSMA/CA~\cite{gklp-ncmcm-18}, requiring receivers to be always on, or 802.15.4e TSCH~\cite{habsw-tschi-16}, allowing for intermittent device sleep.

NDN over LoRa was introduced in \cite{ks-nrclr-17, lndd-endna-20} which required permanent powering of the nodes, depleting the battery. Unfortunately, latency analyses have not been considered. Recent work~\cite{lndd-endna-20} shows the need for a MAC protocol due to high collisions even when deploying only few LoRa nodes.

A system design for ICN over DSME-LoRa is proposed in \cite{kaksw-liiel-22}. 
Based on simulations, the authors find latencies at the order of tens or hundreds of seconds.
In this paper, we close the gap and present a solution to handle these high delays and thus enable common, inter-network IoT~deployments.

\paragraph{Delay-tolerant ICN}
Another ICN~application domain that is challenged by long delays are satellite networks.
Siris~\etal~\cite{svpl-inais-12} find that hop-wise transfer and caching help to increase performance in such networks. They consider an Interest as a long-lived subscription.
In contrast, Kumari~\etal~\cite{ku-ndnii-19} argue that NDN is not viable in satellite scenarios, due to inefficient polling. This is in line with our experimental results.
To reduce long delays and needless retransmissions during satellite handovers, the adjustment of the forwarding path is proposed~\cite{lxtzz-allsc-21}. This solution requires a signal after connecting to a new satellite.

Carofiglio~\etal~\cite{cmprz-liicl-16} exploit link signaling to indicate some kind of loss to trigger a PIT lookup and eventual retransmits, reducing RTTs and redundant retransmits.
LoRa lacks such signaling capabilities.
We incorporate link awareness in our proposed DSME-LoRa gateway.

Kuai~\etal~\cite{khy-dfsnd-19} propose delay-tolerant NDN forwarding for vehicular networks. Fundamentally, neighbored nodes overhear surrounding traffic and adjust their retransmission procedure based on directional network density. 
In simulations, the authors assume a relatively high PIT timeout of 50\,s.
To prevent large PIT tables due to unnecessary long-lasting entries, NACK data packets can include instructions when to retransmit an Interest~\cite{mwz-cpcal-15,ccgt-nnnna-15}.
Similarly to delay-tolerant networking with NDN~\cite{lkz-sdtnc-20}, the IoT requires a mechanism apart from pure request-response.

\paragraph{Producer-initiated ICN}
Burke~\etal~\cite{bgnt-ssndn-14} propose push-based sensor data dissemination, accepting names within a distinct namespace on the consumer.
Gündogan~\etal~\cite{gksw-hrrpi-18} evaluate name indication that triggers a conventional Interest request on the consumer.
Kr\'{o}l~\etal~\cite{khokp-rrmii-18} introduce a nested 4-way handshake to enable RMI use cases based on ICN principles, and analyze drawbacks from long latencies.
This approach is in line with \textit{reflexive forwarding}~\cite{draft-oran-icnrg-reflexive-forwarding}.
We exploit both push and indication concepts in our evaluation.

\section{Conclusions and Future Work}\label{sec:conclusions}
Interconnecting networks with vastly different RTTs is challenging for
any non-trivial communication system, including ICN. ICN, unlike other
frameworks, however, has the unique potential to enable robust
communication to nodes in challenged edge networks without requiring
application layer relays.  In conjunction with an OS-level
implementation of ICN (and extensions), DSME, and LoRa, our two
protocol mechanisms for Internet consumer-initiated and LoRa
producer-initiated communication exhibit high reliability and targeted
completion time (compared to Vanilla ICN) when applied to the
delay-prone regime.  Despite an additional round trip, our evaluations
show low overhead of these approaches, by overcoming redundant
polling.  We leveraged recently proposed gateway behavior (like RICE)
and ICN protocol extensions ({\em reflexive forwarding}), the latter
of which serves many other use cases beyond {\em phoning home} and
could be considered a useful standard ICN feature.

This work leads to interesting future research:
First, we will
integrate an estimator model in the gateway, aiming to reduce the RTT
in our {\em Delay-tolerant retrieval} case. This relieves consumer knowledge, \eg to estimate domain specific retry timers individually.
Second, we will explore security aspects.
This includes, but is not limited to,
bootstrapping of LoRa nodes and gateways,
a secure registration process which requires trust to the gateway,
and authentication of a LoRa node before the gateway acts on its behalf.
We will further derive a threat model for end-to-end consumer-to-producer security.
Third, we will evaluate the scalability and robustness of our ICN
protocol framework in more complex topologies (multi gateway, node to
node) to demonstrate data sharing benefits.
Finally, we want to
investigate additional use cases, including Remote Method Invocation
on LoRa nodes and multicast-style communication, \eg for distributing
firmware updates to LoRa nodes.

\newpage

\label{lastbodypage}

\balance
\bibliographystyle{ACM-Reference-Format}
\bibliography{own,rfcs,ids,ngi,iot,layer2,meta,complexity,internet}


\begin{thebibliography}{61}


\ifx \showCODEN    \undefined \def \showCODEN     #1{\unskip}     \fi
\ifx \showDOI      \undefined \def \showDOI       #1{#1}\fi
\ifx \showISBNx    \undefined \def \showISBNx     #1{\unskip}     \fi
\ifx \showISBNxiii \undefined \def \showISBNxiii  #1{\unskip}     \fi
\ifx \showISSN     \undefined \def \showISSN      #1{\unskip}     \fi
\ifx \showLCCN     \undefined \def \showLCCN      #1{\unskip}     \fi
\ifx \shownote     \undefined \def \shownote      #1{#1}          \fi
\ifx \showarticletitle \undefined \def \showarticletitle #1{#1}   \fi
\ifx \showURL      \undefined \def \showURL       {\relax}        \fi
\providecommand\bibfield[2]{#2}
\providecommand\bibinfo[2]{#2}
\providecommand\natexlab[1]{#1}
\providecommand\showeprint[2][]{arXiv:#2}

\bibitem[Afanasyev et~al\mbox{.}(2021)]%
        {aszzm-ndg-21}
\bibfield{author}{\bibinfo{person}{Alexander Afanasyev},
  \bibinfo{person}{Junxiao Shi}, \bibinfo{person}{Beichuan Zhang},
  \bibinfo{person}{Lixia Zhang}, \bibinfo{person}{Ilya Moiseenko},
  \bibinfo{person}{Yingdi Yu}, \bibinfo{person}{Wentao Shang},
  \bibinfo{person}{Yanbiao Li}, \bibinfo{person}{Spyridon Mastorakis},
  \bibinfo{person}{Yi Huang}, \bibinfo{person}{Jerald~Paul Abraham},
  \bibinfo{person}{Eric Newberry}, \bibinfo{person}{Steve DiBenedetto},
  \bibinfo{person}{Chengyu Fan}, \bibinfo{person}{Christos Papadopoulos},
  \bibinfo{person}{Davide Pesavento}, \bibinfo{person}{Giulio Grassi},
  \bibinfo{person}{Giovanni Pau}, \bibinfo{person}{Hang Zhang},
  \bibinfo{person}{Tian Song}, \bibinfo{person}{Haowei Yuan},
  \bibinfo{person}{Hila~Ben Abraham}, \bibinfo{person}{Patrick Crowley},
  \bibinfo{person}{Syed~Obaid Amin}, \bibinfo{person}{Vince Lehman},
  \bibinfo{person}{Muktadir Chowdhury}, \bibinfo{person}{}, {and}
  \bibinfo{person}{Lan Wang}.} \bibinfo{year}{2021}\natexlab{}.
\newblock \bibinfo{booktitle}{\emph{{NFD Developer's Guide}}}.
\newblock \bibinfo{type}{Technical Report} NDN-0021.
  \bibinfo{institution}{NDN}.
\newblock
\urldef\tempurl%
\url{https://named-data.net/publications/techreports/ndn-0021-11-nfd-guide/}
\showURL{%
\tempurl}


\bibitem[Alamos et~al\mbox{.}(2022a)]%
        {aksw-dslrc-22}
\bibfield{author}{\bibinfo{person}{Jose Alamos}, \bibinfo{person}{Peter
  Kietzmann}, \bibinfo{person}{Thomas~C. Schmidt}, {and}
  \bibinfo{person}{Matthias W{\"a}hlisch}.} \bibinfo{year}{2022}\natexlab{a}.
\newblock \showarticletitle{{DSME-LoRa: Seamless Long Range Communication
  Between Arbitrary Nodes in the Constrained IoT}}.
\newblock \bibinfo{journal}{\emph{Transactions on Sensor Networks (TOSN)}}
  (\bibinfo{year}{2022}).
\newblock
\urldef\tempurl%
\url{https://dl.acm.org/doi/10.1145/3552432}
\showURL{%
\tempurl}


\bibitem[Alamos et~al\mbox{.}(2022b)]%
        {aksw-edmls-22}
\bibfield{author}{\bibinfo{person}{Jose Alamos}, \bibinfo{person}{Peter
  Kietzmann}, \bibinfo{person}{Thomas~C. Schmidt}, {and}
  \bibinfo{person}{Matthias W{\"a}hlisch}.} \bibinfo{year}{2022}\natexlab{b}.
\newblock \showarticletitle{{WIP: Exploring DSME MAC for LoRa -- A System
  Integration and First Evaluation}}. In \bibinfo{booktitle}{\emph{23rd IEEE
  International Symposium on a World of Wireless, Mobile and Multimedia
  Networks (WoWMoM)}} (Belfast, UK). \bibinfo{publisher}{IEEE},
  \bibinfo{address}{Piscataway, NJ, USA}.
\newblock
\urldef\tempurl%
\url{https://doi.org/10.1109/WoWMoM54355.2022.00050}
\showURL{%
\tempurl}


\bibitem[Amadeo et~al\mbox{.}(2014)]%
        {acim-ndnia-14}
\bibfield{author}{\bibinfo{person}{M. Amadeo}, \bibinfo{person}{C. Campolo},
  \bibinfo{person}{A. Iera}, {and} \bibinfo{person}{A. Molinaro}.}
  \bibinfo{year}{2014}\natexlab{}.
\newblock \showarticletitle{{Named data networking for IoT: An architectural
  perspective}}. In \bibinfo{booktitle}{\emph{2014 European Conference on
  Networks and Communications (EuCNC)}}. \bibinfo{publisher}{IEEE},
  \bibinfo{address}{Piscataway, NJ, USA}, \bibinfo{pages}{1--5}.
\newblock


\bibitem[Arshad et~al\mbox{.}(2019)]%
        {aarl-raini-19}
\bibfield{author}{\bibinfo{person}{Sobia Arshad},
  \bibinfo{person}{Muhammad~Awais Azam}, \bibinfo{person}{Mubashir~Husain
  Rehmani}, {and} \bibinfo{person}{Jonathan Loo}.}
  \bibinfo{year}{2019}\natexlab{}.
\newblock \showarticletitle{{Recent Advances in Information-Centric
  Networking-Based Internet of Things (ICN-IoT)}}.
\newblock \bibinfo{journal}{\emph{IEEE Internet of Things Journal}}
  \bibinfo{volume}{6}, \bibinfo{number}{2} (\bibinfo{year}{2019}),
  \bibinfo{pages}{2128--2158}.
\newblock


\bibitem[Baccelli et~al\mbox{.}(2018)]%
        {bghkl-rosos-18}
\bibfield{author}{\bibinfo{person}{Emmanuel Baccelli}, \bibinfo{person}{Cenk
  G{\"u}ndogan}, \bibinfo{person}{Oliver Hahm}, \bibinfo{person}{Peter
  Kietzmann}, \bibinfo{person}{Martine Lenders}, \bibinfo{person}{Hauke
  Petersen}, \bibinfo{person}{Kaspar Schleiser}, \bibinfo{person}{Thomas~C.
  Schmidt}, {and} \bibinfo{person}{Matthias W{\"a}hlisch}.}
  \bibinfo{year}{2018}\natexlab{}.
\newblock \showarticletitle{{RIOT: an Open Source Operating System for Low-end
  Embedded Devices in the IoT}}.
\newblock \bibinfo{journal}{\emph{IEEE Internet of Things Journal}}
  \bibinfo{volume}{5}, \bibinfo{number}{6} (\bibinfo{date}{December}
  \bibinfo{year}{2018}), \bibinfo{pages}{4428--4440}.
\newblock
\urldef\tempurl%
\url{http://dx.doi.org/10.1109/JIOT.2018.2815038}
\showURL{%
\tempurl}


\bibitem[Baccelli et~al\mbox{.}(2014)]%
        {bmhsw-icnie-14}
\bibfield{author}{\bibinfo{person}{Emmanuel Baccelli},
  \bibinfo{person}{Christian Mehlis}, \bibinfo{person}{Oliver Hahm},
  \bibinfo{person}{Thomas~C. Schmidt}, {and} \bibinfo{person}{Matthias
  W{\"a}hlisch}.} \bibinfo{year}{2014}\natexlab{}.
\newblock \showarticletitle{{Information Centric Networking in the IoT:
  Experiments with NDN in the Wild}}. In \bibinfo{booktitle}{\emph{Proc. of 1st
  ACM Conf. on Information-Centric Networking (ICN-2014)}} (Paris).
  \bibinfo{publisher}{ACM}, \bibinfo{address}{New York},
  \bibinfo{pages}{77--86}.
\newblock
\urldef\tempurl%
\url{http://dx.doi.org/10.1145/2660129.2660144}
\showURL{%
\tempurl}


\bibitem[Bezunartea et~al\mbox{.}(2019)]%
        {bgbts-teelm-19}
\bibfield{author}{\bibinfo{person}{Maite Bezunartea},
  \bibinfo{person}{Roald~Van Glabbeek}, \bibinfo{person}{An Braeken},
  \bibinfo{person}{Jacques Tiberghien}, {and} \bibinfo{person}{Kris
  Steenhaut}.} \bibinfo{year}{2019}\natexlab{}.
\newblock \showarticletitle{{Towards Energy Efficient LoRa Multihop Networks}}.
  In \bibinfo{booktitle}{\emph{International Symposium on Local and
  Metropolitan Area Networks (LANMAN '19)}} (Paris, France).
  \bibinfo{publisher}{IEEE}, \bibinfo{address}{Piscataway, NJ, USA},
  \bibinfo{pages}{1--3}.
\newblock


\bibitem[Burke et~al\mbox{.}(2014)]%
        {bgnt-ssndn-14}
\bibfield{author}{\bibinfo{person}{Jeff Burke}, \bibinfo{person}{Paolo Gasti},
  \bibinfo{person}{Naveen Nathan}, {and} \bibinfo{person}{Gene Tsudik}.}
  \bibinfo{year}{2014}\natexlab{}.
\newblock \showarticletitle{{Secure Sensing over Named Data Networking}}. In
  \bibinfo{booktitle}{\emph{13th International Symposium on Network Computing
  and Applications (NCA'14)}}. \bibinfo{publisher}{IEEE Computer Society},
  \bibinfo{address}{Washington, DC, USA}, \bibinfo{pages}{175--180}.
\newblock


\bibitem[Carofiglio et~al\mbox{.}(2016)]%
        {cmprz-liicl-16}
\bibfield{author}{\bibinfo{person}{Giovanna Carofiglio}, \bibinfo{person}{Luca
  Muscariello}, \bibinfo{person}{Michele Papalini}, \bibinfo{person}{Natalya
  Rozhnova}, {and} \bibinfo{person}{Xuan Zeng}.}
  \bibinfo{year}{2016}\natexlab{}.
\newblock \showarticletitle{{Leveraging ICN In-Network Control for Loss
  Detection and Recovery in Wireless Mobile Networks}}. In
  \bibinfo{booktitle}{\emph{Proceedings of the 3rd ACM Conference on
  Information-Centric Networking}}. \bibinfo{publisher}{ACM},
  \bibinfo{address}{New York, NY, USA}, \bibinfo{pages}{50--59}.
\newblock


\bibitem[Carzaniga et~al\mbox{.}(2011)]%
        {cpw-cpsni-11}
\bibfield{author}{\bibinfo{person}{Antonio Carzaniga}, \bibinfo{person}{Michele
  Papalini}, {and} \bibinfo{person}{Alexander~L. Wolf}.}
  \bibinfo{year}{2011}\natexlab{}.
\newblock \showarticletitle{{Content-based Publish/Subscribe Networking and
  Information-centric Networking}}. In \bibinfo{booktitle}{\emph{Proc. of the
  ACM SIGCOMM WS on Information-centric Networking (ICN '11)}} (Toronto,
  Ontario, Canada). \bibinfo{publisher}{ACM}, \bibinfo{address}{New York, NY,
  USA}, \bibinfo{pages}{56--61}.
\newblock


\bibitem[Compagno et~al\mbox{.}(2015)]%
        {ccgt-nnnna-15}
\bibfield{author}{\bibinfo{person}{Alberto Compagno}, \bibinfo{person}{Mauro
  Conti}, \bibinfo{person}{Cesar Ghali}, {and} \bibinfo{person}{Gene Tsudik}.}
  \bibinfo{year}{2015}\natexlab{}.
\newblock \showarticletitle{{To NACK or Not to NACK? Negative Acknowledgments
  in Information-Centric Networking}}. In \bibinfo{booktitle}{\emph{24th
  International Conference on Computer Communication and Networks (ICCCN'15)}}.
  \bibinfo{publisher}{IEEE Press}, \bibinfo{address}{Piscataway, NJ, USA},
  \bibinfo{numpages}{10}~pages.
\newblock


\bibitem[Cotrim and {a}o Henrique~Kleinschmidt(2020)]%
        {ck-lmnrc-20}
\bibfield{author}{\bibinfo{person}{Jeferson~Rodrigues Cotrim} {and}
  \bibinfo{person}{Jo\ {a}o Henrique~Kleinschmidt}.}
  \bibinfo{year}{2020}\natexlab{}.
\newblock \showarticletitle{{LoRaWAN Mesh Networks: A Review and Classification
  of Multihop Communication}}.
\newblock \bibinfo{journal}{\emph{Sensors}} \bibinfo{volume}{20},
  \bibinfo{number}{15} (\bibinfo{year}{2020}), \bibinfo{pages}{4273}.
\newblock


\bibitem[Elbsir et~al\mbox{.}(2020a)]%
        {ekbb-elcbe-20}
\bibfield{author}{\bibinfo{person}{Houssem~Eddin Elbsir},
  \bibinfo{person}{Mohammed Kassab}, \bibinfo{person}{Sami Bhiri}, {and}
  \bibinfo{person}{Mohamed~Hedi Bedoui}.} \bibinfo{year}{2020}\natexlab{a}.
\newblock \showarticletitle{{Evaluation of LoRaWAN Class B efficiency for
  downlink traffic}}. In \bibinfo{booktitle}{\emph{2020 16th International
  Conference on Wireless and Mobile Computing, Networking and Communications
  (WiMob)}} (Thessaloniki, Greece). \bibinfo{publisher}{IEEE},
  \bibinfo{address}{Piscataway, NJ, USA}, \bibinfo{pages}{105--110}.
\newblock


\bibitem[Elbsir et~al\mbox{.}(2020b)]%
        {ekbd-elced-20}
\bibfield{author}{\bibinfo{person}{Houssem~Eddin Elbsir},
  \bibinfo{person}{Mohammed Kassab}, \bibinfo{person}{Sami Bhiri}, {and}
  \bibinfo{person}{Mohamed~Hedi Bedoui}.} \bibinfo{year}{2020}\natexlab{b}.
\newblock \showarticletitle{{Evaluation of LoRaWAN Class B efficiency for
  downlink traffic}}. In \bibinfo{booktitle}{\emph{16th International
  Conference on Wireless and Mobile Computing, Networking and Communications
  (WiMob'20)}}. \bibinfo{publisher}{IEEE}, \bibinfo{address}{Piscataway, NJ,
  USA}, \bibinfo{pages}{105--110}.
\newblock


\bibitem[Ferre(2017)]%
        {f-cplal-17}
\bibfield{author}{\bibinfo{person}{Guillaume Ferre}.}
  \bibinfo{year}{2017}\natexlab{}.
\newblock \showarticletitle{{Collision and packet loss analysis in a LoRaWAN
  network}}. In \bibinfo{booktitle}{\emph{25th European Signal Processing
  Conference (EUSIPCO'17)}}. \bibinfo{publisher}{IEEE},
  \bibinfo{address}{Piscataway, NJ, USA}, \bibinfo{pages}{2586--2590}.
\newblock


\bibitem[Finnegan et~al\mbox{.}(2018)]%
        {fbf-eslgc-18}
\bibfield{author}{\bibinfo{person}{Joseph Finnegan}, \bibinfo{person}{Stephen
  Brown}, {and} \bibinfo{person}{Ronan Farrell}.}
  \bibinfo{year}{2018}\natexlab{}.
\newblock \showarticletitle{{Evaluating the Scalability of LoRaWAN Gateways for
  Class B Communication in ns-3}}. In \bibinfo{booktitle}{\emph{IEEE Conference
  on Standards for Communications and Networking (CSCN'18)}}.
  \bibinfo{publisher}{IEEE}, \bibinfo{address}{Piscataway, NJ, USA},
  \bibinfo{pages}{1--6}.
\newblock


\bibitem[Gimenez and Petrov(2021)]%
        {RFC-9011}
\bibfield{author}{\bibinfo{person}{O. Gimenez} {and} \bibinfo{person}{I.
  Petrov}.} \bibinfo{year}{2021}\natexlab{}.
\newblock \bibinfo{booktitle}{\emph{{Static Context Header Compression and
  Fragmentation (SCHC) over LoRaWAN}}}.
\newblock \bibinfo{type}{RFC} 9011. \bibinfo{institution}{IETF}.
\newblock


\bibitem[Gonzalez et~al\mbox{.}(2018)]%
        {gbv-slpld-18}
\bibfield{author}{\bibinfo{person}{Nicolas Gonzalez}, \bibinfo{person}{Adrien
  Van~Den Bossche}, {and} \bibinfo{person}{Thierry Val}.}
  \bibinfo{year}{2018}\natexlab{}.
\newblock \showarticletitle{{Specificities of the LoRa physical layer for the
  development of new ad hoc MAC layers}}. In \bibinfo{booktitle}{\emph{{17th
  International Conference on Ad Hoc Networks and Wireless (AdHoc-Now'18)}}}
  (St Malo, France), Vol.~\bibinfo{volume}{11104}.
  \bibinfo{publisher}{Springer}, \bibinfo{address}{Cham, Switzerland},
  \bibinfo{pages}{163--174}.
\newblock


\bibitem[G{\"u}ndogan et~al\mbox{.}(2018a)]%
        {gklp-ncmcm-18}
\bibfield{author}{\bibinfo{person}{Cenk G{\"u}ndogan}, \bibinfo{person}{Peter
  Kietzmann}, \bibinfo{person}{Martine Lenders}, \bibinfo{person}{Hauke
  Petersen}, \bibinfo{person}{Thomas~C. Schmidt}, {and}
  \bibinfo{person}{Matthias W{\"a}hlisch}.} \bibinfo{year}{2018}\natexlab{a}.
\newblock \showarticletitle{{NDN, CoAP, and MQTT: A Comparative Measurement
  Study in the IoT}}. In \bibinfo{booktitle}{\emph{Proc. of 5th ACM Conference
  on Information-Centric Networking (ICN)}}. \bibinfo{publisher}{ACM},
  \bibinfo{address}{New York, NY, USA}, \bibinfo{pages}{159--171}.
\newblock
\urldef\tempurl%
\url{https://doi.org/10.1145/3267955.3267967}
\showURL{%
\tempurl}


\bibitem[G{\"u}ndogan et~al\mbox{.}(2018b)]%
        {gksw-hrrpi-18}
\bibfield{author}{\bibinfo{person}{Cenk G{\"u}ndogan}, \bibinfo{person}{Peter
  Kietzmann}, \bibinfo{person}{Thomas~C. Schmidt}, {and}
  \bibinfo{person}{Matthias W{\"a}hlisch}.} \bibinfo{year}{2018}\natexlab{b}.
\newblock \showarticletitle{{HoPP: Robust and Resilient Publish-Subscribe for
  an Information-Centric Internet of Things}}. In
  \bibinfo{booktitle}{\emph{Proc. of the 43rd IEEE Conference on Local Computer
  Networks (LCN)}} (Chicago, USA). \bibinfo{publisher}{IEEE Press},
  \bibinfo{address}{Piscataway, NJ, USA}, \bibinfo{pages}{331--334}.
\newblock
\urldef\tempurl%
\url{http://doi.org/10.1109/LCN.2018.8638030}
\showURL{%
\tempurl}


\bibitem[G{\"u}ndogan et~al\mbox{.}(2019)]%
        {gksw-innlp-19}
\bibfield{author}{\bibinfo{person}{Cenk G{\"u}ndogan}, \bibinfo{person}{Peter
  Kietzmann}, \bibinfo{person}{Thomas~C. Schmidt}, {and}
  \bibinfo{person}{Matthias W{\"a}hlisch}.} \bibinfo{year}{2019}\natexlab{}.
\newblock \showarticletitle{{ICNLoWPAN -- Named-Data Networking in Low Power
  IoT Networks}}. In \bibinfo{booktitle}{\emph{Proc. of 18th IFIP Networking
  Conference}} (Warsaw, Poland). \bibinfo{publisher}{IEEE Press},
  \bibinfo{address}{Piscataway, NJ, USA}, \bibinfo{pages}{1--9}.
\newblock
\urldef\tempurl%
\url{http://dx.doi.org/10.23919/IFIPNetworking.2019.8816850}
\showURL{%
\tempurl}


\bibitem[Hahm et~al\mbox{.}(2016)]%
        {habsw-tschi-16}
\bibfield{author}{\bibinfo{person}{Oliver Hahm}, \bibinfo{person}{Cedric
  Adjih}, \bibinfo{person}{Emmanuel Baccelli}, \bibinfo{person}{Thomas~C.
  Schmidt}, {and} \bibinfo{person}{Matthias W{\"a}hlisch}.}
  \bibinfo{year}{2016}\natexlab{}.
\newblock \showarticletitle{{Designing Time Slotted Channel Hopping and
  {Information-Centric} Networking for {IoT}}}. In
  \bibinfo{booktitle}{\emph{Proc. of 8th IFIP International Conference on New
  Technologies, Mobility \& Security (NTMS)}} (Larnaca, Cyprus).
  \bibinfo{publisher}{IEEE Press}, \bibinfo{address}{Piscataway, NJ, USA},
  \bibinfo{numpages}{5}~pages.
\newblock


\bibitem[Haubro et~al\mbox{.}(2020)]%
        {hoof-tlrri-20}
\bibfield{author}{\bibinfo{person}{Martin Haubro}, \bibinfo{person}{Charalampos
  Orfanidis}, \bibinfo{person}{George Oikonomou}, {and}
  \bibinfo{person}{Xenofon Fafoutis}.} \bibinfo{year}{2020}\natexlab{}.
\newblock \showarticletitle{{TSCH-over-LoRA: long range and reliable IPv6
  multi-hop networks for the internet of things}}.
\newblock \bibinfo{journal}{\emph{Internet Technology Letters}}
  \bibinfo{volume}{3}, \bibinfo{number}{4} (\bibinfo{year}{2020}),
  \bibinfo{pages}{e165}.
\newblock


\bibitem[{IEEE 802.15 Working Group}(2016)]%
        {IEEE-802.15.4-16}
\bibfield{author}{\bibinfo{person}{{IEEE 802.15 Working Group}}.}
  \bibinfo{year}{2016}\natexlab{}.
\newblock \bibinfo{booktitle}{\emph{{IEEE Standard for Low-Rate Wireless
  Networks}}}.
\newblock \bibinfo{type}{{T}echnical {R}eport} IEEE Std
  802.15.4\texttrademark--2015 (Revision of IEEE Std 802.15.4-2011).
  \bibinfo{institution}{IEEE}, \bibinfo{address}{New York, NY, USA}.
  \bibinfo{pages}{1--709} pages.
\newblock


\bibitem[Karunathilake et~al\mbox{.}(2021)]%
        {kuf-ldlif-21}
\bibfield{author}{\bibinfo{person}{Thenuka Karunathilake},
  \bibinfo{person}{Asanga Udugama}, {and} \bibinfo{person}{Anna
  F{\"{o}}rster}.} \bibinfo{year}{2021}\natexlab{}.
\newblock \showarticletitle{{LoRa-DuCy: Duty Cycling for LoRa-Enabled Internet
  of Things Devices}}. In \bibinfo{booktitle}{\emph{12th International
  Conference on Ubiquitous and Future Networks (ICUFN '21)}}.
  \bibinfo{publisher}{IEEE}, \bibinfo{address}{Piscataway, NJ, USA},
  \bibinfo{pages}{283--288}.
\newblock


\bibitem[Kauer et~al\mbox{.}(2018)]%
        {kkt-rwmnd-18}
\bibfield{author}{\bibinfo{person}{Florian Kauer}, \bibinfo{person}{Maximilian
  K{\"{o}}stler}, {and} \bibinfo{person}{Volker Turau}.}
  \bibinfo{year}{2018}\natexlab{}.
\newblock \bibinfo{booktitle}{\emph{{Reliable Wireless Multi-Hop Networks with
  Decentralized Slot Management: An Analysis of IEEE 802.15.4 DSME}}}.
\newblock \bibinfo{type}{Technical Report} arXiv:1806.10521.
  \bibinfo{institution}{Open Archive: arXiv.org}.
\newblock


\bibitem[Kietzmann et~al\mbox{.}(2022)]%
        {kaksw-liiel-22}
\bibfield{author}{\bibinfo{person}{Peter Kietzmann}, \bibinfo{person}{Jose
  Alamos}, \bibinfo{person}{Dirk Kutscher}, \bibinfo{person}{Thomas~C.
  Schmidt}, {and} \bibinfo{person}{Matthias W{\"a}hlisch}.}
  \bibinfo{year}{2022}\natexlab{}.
\newblock \showarticletitle{{Long-Range ICN for the IoT: Exploring a LoRa
  System Design}}. In \bibinfo{booktitle}{\emph{Proc. of 21th IFIP Networking
  Conference}} (Catania, Italy). \bibinfo{publisher}{IEEE Press},
  \bibinfo{address}{Piscataway, NJ, USA}.
\newblock
\urldef\tempurl%
\url{https://doi.org/10.23919/IFIPNetworking55013.2022.9829792}
\showURL{%
\tempurl}


\bibitem[Kietzmann et~al\mbox{.}(2017)]%
        {kgshw-nnmam-17}
\bibfield{author}{\bibinfo{person}{Peter Kietzmann}, \bibinfo{person}{Cenk
  G{\"u}ndogan}, \bibinfo{person}{Thomas~C. Schmidt}, \bibinfo{person}{Oliver
  Hahm}, {and} \bibinfo{person}{Matthias W{\"a}hlisch}.}
  \bibinfo{year}{2017}\natexlab{}.
\newblock \showarticletitle{{The Need for a Name to MAC Address Mapping in NDN:
  Towards Quantifying the Resource Gain}}. In \bibinfo{booktitle}{\emph{Proc.
  of 4th ACM Conference on Information-Centric Networking (ICN)}}.
  \bibinfo{publisher}{ACM}, \bibinfo{address}{New York, NY, USA},
  \bibinfo{pages}{36--42}.
\newblock
\urldef\tempurl%
\url{https://dl.acm.org/doi/10.1145/3125719.3125737}
\showURL{%
\tempurl}


\bibitem[King and Sugiyama(2017)]%
        {ks-nrclr-17}
\bibfield{author}{\bibinfo{person}{Randy King} {and} \bibinfo{person}{Rod
  Sugiyama}.} \bibinfo{year}{2017}\natexlab{}.
\newblock \bibinfo{booktitle}{\emph{{A New Remote Communications Link to Reduce
  Residential PV Solar Costs}}}.
\newblock \bibinfo{type}{{T}echnical {R}eport} DE-EE0007592.
  \bibinfo{institution}{U.S. Department of Energy Office of Scientific and
  Technical Information}.
\newblock


\bibitem[Kr\'{o}l et~al\mbox{.}(2018)]%
        {khokp-rrmii-18}
\bibfield{author}{\bibinfo{person}{Micha\l{} Kr\'{o}l}, \bibinfo{person}{Karim
  Habak}, \bibinfo{person}{David Oran}, \bibinfo{person}{Dirk Kutscher}, {and}
  \bibinfo{person}{Ioannis Psaras}.} \bibinfo{year}{2018}\natexlab{}.
\newblock \showarticletitle{{RICE: Remote Method Invocation in ICN}}. In
  \bibinfo{booktitle}{\emph{Proceedings of the 5th ACM Conference on
  Information-Centric Networking}} (Boston, Massachusetts)
  \emph{(\bibinfo{series}{ICN '18})}. \bibinfo{publisher}{ACM},
  \bibinfo{address}{New York, NY, USA}, \bibinfo{pages}{1--11}.
\newblock
\showISBNx{9781450359597}


\bibitem[Kuai et~al\mbox{.}(2019)]%
        {khy-dfsnd-19}
\bibfield{author}{\bibinfo{person}{Meng Kuai}, \bibinfo{person}{Xiaoyan Hong},
  {and} \bibinfo{person}{Qiangyuan Yu}.} \bibinfo{year}{2019}\natexlab{}.
\newblock \showarticletitle{{Delay-tolerant forwarding strategy for named data
  networking in vehicular environment}}.
\newblock \bibinfo{journal}{\emph{International Journal of Ad Hoc and
  Ubiquitous Computing}} \bibinfo{volume}{31}, \bibinfo{number}{1}
  (\bibinfo{year}{2019}), \bibinfo{numpages}{12}~pages.
\newblock


\bibitem[Kumari and Ujjwal(2019)]%
        {ku-ndnii-19}
\bibfield{author}{\bibinfo{person}{Ritika Kumari} {and} \bibinfo{person}{R.~L.
  Ujjwal}.} \bibinfo{year}{2019}\natexlab{}.
\newblock \showarticletitle{{Name Data Networking for Interplanetary Internet:
  An Architectural Perspective}}.
\newblock \bibinfo{journal}{\emph{International Journal of Research in Advent
  Technology}} \bibinfo{volume}{7}, \bibinfo{number}{5} (\bibinfo{year}{2019}),
  \bibinfo{pages}{436--441}.
\newblock


\bibitem[Lenders et~al\mbox{.}(2020)]%
        {lgsw-cdsfr-20}
\bibfield{author}{\bibinfo{person}{Martine~Sophie Lenders},
  \bibinfo{person}{Cenk G{\"u}ndogan}, \bibinfo{person}{Thomas~C. Schmidt},
  {and} \bibinfo{person}{Matthias W{\"a}hlisch}.}
  \bibinfo{year}{2020}\natexlab{}.
\newblock \showarticletitle{{Connecting the Dots: Selective Fragment Recovery
  in ICNLoWPAN}}. In \bibinfo{booktitle}{\emph{Proc. of 7th ACM Conference on
  Information-Centric Networking (ICN)}} (Montreal, CA).
  \bibinfo{publisher}{ACM}, \bibinfo{address}{New York},
  \bibinfo{pages}{70--76}.
\newblock
\urldef\tempurl%
\url{https://doi.org/10.1145/3405656.3418719}
\showURL{%
\tempurl}


\bibitem[Leonardi et~al\mbox{.}(2020)]%
        {lbbp-calma-20}
\bibfield{author}{\bibinfo{person}{Luca Leonardi}, \bibinfo{person}{Lucia~Lo
  Bello}, \bibinfo{person}{Filippo Battaglia}, {and} \bibinfo{person}{Gaetano
  Patti}.} \bibinfo{year}{2020}\natexlab{}.
\newblock \showarticletitle{{Comparative Assessment of the LoRaWAN Medium
  Access Control Protocols for IoT: Does Listen before Talk Perform Better than
  ALOHA?}}
\newblock \bibinfo{journal}{\emph{Electronics}} \bibinfo{volume}{9},
  \bibinfo{number}{4} (\bibinfo{year}{2020}), \bibinfo{pages}{553}.
\newblock


\bibitem[Li et~al\mbox{.}(2020)]%
        {lkz-sdtnc-20}
\bibfield{author}{\bibinfo{person}{Tianxiang Li}, \bibinfo{person}{Zhaoning
  Kong}, {and} \bibinfo{person}{Lixia Zhang}.} \bibinfo{year}{2020}\natexlab{}.
\newblock \showarticletitle{{Supporting Delay Tolerant Networking: A
  Comparative Study of Epidemic Routing and NDN}}. In
  \bibinfo{booktitle}{\emph{International Conference on Communications
  Workshops (ICC'20 Workshop)}}. \bibinfo{publisher}{IEEE Press},
  \bibinfo{address}{Piscataway, NJ, USA}, \bibinfo{numpages}{6}~pages.
\newblock


\bibitem[Liang et~al\mbox{.}(2021)]%
        {lxtzz-allsc-21}
\bibfield{author}{\bibinfo{person}{Teng Liang}, \bibinfo{person}{Zhongda Xia},
  \bibinfo{person}{Guoming Tang}, \bibinfo{person}{Yu Zhang}, {and}
  \bibinfo{person}{Beichuan Zhang}.} \bibinfo{year}{2021}\natexlab{}.
\newblock \showarticletitle{{NDN in Large LEO Satellite Constellations: A Case
  of Consumer Mobility Support}}. In \bibinfo{booktitle}{\emph{Proceedings of
  the 8th ACM Conference on Information-Centric Networking}}.
  \bibinfo{publisher}{ACM}, \bibinfo{address}{New York, NY, USA},
  \bibinfo{numpages}{12}~pages.
\newblock


\bibitem[Liu et~al\mbox{.}(2020)]%
        {lndd-endna-20}
\bibfield{author}{\bibinfo{person}{Yaoqing Liu}, \bibinfo{person}{Laurent
  Njilla}, \bibinfo{person}{Anthony Dowling}, {and} \bibinfo{person}{Wan Du}.}
  \bibinfo{year}{2020}\natexlab{}.
\newblock \showarticletitle{Empowering Named Data Networks for Ad-Hoc
  Long-Range Communication}. In \bibinfo{booktitle}{\emph{Wireless and Optical
  Communications Conference (WOCC'20)}}. \bibinfo{publisher}{IEEE},
  \bibinfo{address}{Piscataway, NJ, USA}, \bibinfo{pages}{1--6}.
\newblock


\bibitem[{LoRa Alliance -- Technical Committee}(2017)]%
        {lorawan-spec-11}
\bibfield{author}{\bibinfo{person}{{LoRa Alliance -- Technical Committee}}.}
  \bibinfo{year}{2017}\natexlab{}.
\newblock \bibinfo{booktitle}{\emph{LoRaWAN 1.1 Specification}}.
\newblock \bibinfo{type}{{T}echnical {R}eport}. \bibinfo{institution}{LoRa
  Alliance}.
\newblock
\urldef\tempurl%
\url{https://lora-alliance.org/sites/default/files/2018-04/lorawantm_specification_-v1.1.pdf}
\showURL{%
\tempurl}


\bibitem[Mathieu et~al\mbox{.}(2016)]%
        {mwt-tucin-16}
\bibfield{author}{\bibinfo{person}{Bertrand Mathieu}, \bibinfo{person}{Cedric
  Westphal}, {and} \bibinfo{person}{Patrick Truong}.}
  \bibinfo{year}{2016}\natexlab{}.
\newblock \showarticletitle{Towards the Usage of CCN for IoT Networks}.
\newblock In \bibinfo{booktitle}{\emph{Internet of Things (IoT) in 5G Mobile
  Technologies}}. \bibinfo{publisher}{Springer}, \bibinfo{address}{Cham,
  Switzerland}, \bibinfo{pages}{3--24}.
\newblock


\bibitem[Mikhaylov et~al\mbox{.}(2018)]%
        {mpp-edtpl-18}
\bibfield{author}{\bibinfo{person}{Konstantin Mikhaylov}, \bibinfo{person}{Juha
  Pet\"{a}j\"{a}j\"{a}rvi}, {and} \bibinfo{person}{Ari Pouttu}.}
  \bibinfo{year}{2018}\natexlab{}.
\newblock \showarticletitle{{Effect of Downlink Traffic on Performance of
  LoRaWAN LPWA Networks: Empirical Study}}. In \bibinfo{booktitle}{\emph{29th
  Annual International Symposium on Personal, Indoor and Mobile Radio
  Communications (PIMRC'18)}}. \bibinfo{publisher}{IEEE},
  \bibinfo{address}{Piscataway, NJ, USA}, \bibinfo{numpages}{6}~pages.
\newblock


\bibitem[{Mininet Project Contributors}(2022)]%
        {mininet-framework-22}
\bibfield{author}{\bibinfo{person}{{Mininet Project Contributors}}.}
  \bibinfo{year}{2022}\natexlab{}.
\newblock \bibinfo{title}{{Mininet - An Instant Virtual Network on your Laptop
  (or other PC)}}.
\newblock \bibinfo{howpublished}{\url{http://www.mininet.org/}, last accessed
  09-06-2022}.
\newblock


\bibitem[Moiseenko et~al\mbox{.}(2015)]%
        {mwz-cpcal-15}
\bibfield{author}{\bibinfo{person}{Ilya Moiseenko}, \bibinfo{person}{Lijing
  Wang}, {and} \bibinfo{person}{Lixia Zhang}.} \bibinfo{year}{2015}\natexlab{}.
\newblock \showarticletitle{{Consumer / Producer Communication with Application
  Level Framing in Named Data Networking}}. In
  \bibinfo{booktitle}{\emph{Proceedings of the 2nd ACM Conference on
  Information-Centric Networking}} (San Francisco, California, USA)
  \emph{(\bibinfo{series}{ICN '15})}. \bibinfo{publisher}{ACM},
  \bibinfo{address}{New York, NY, USA}, \bibinfo{pages}{99--108}.
\newblock
\showISBNx{9781450338554}


\bibitem[Mosko et~al\mbox{.}(2019)]%
        {RFC-8569}
\bibfield{author}{\bibinfo{person}{M. Mosko}, \bibinfo{person}{I. Solis}, {and}
  \bibinfo{person}{C. Wood}.} \bibinfo{year}{2019}\natexlab{}.
\newblock \bibinfo{booktitle}{\emph{{Content-Centric Networking (CCNx)
  Semantics}}}.
\newblock \bibinfo{type}{RFC} 8569. \bibinfo{institution}{IETF}.
\newblock


\bibitem[O'Kennedy et~al\mbox{.}(2020)]%
        {knwm-pecsl-20}
\bibfield{author}{\bibinfo{person}{Morgan O'Kennedy}, \bibinfo{person}{Thomas
  Niesler}, \bibinfo{person}{Riaan Wolhuter}, {and} \bibinfo{person}{Nathalie
  Mitton}.} \bibinfo{year}{2020}\natexlab{}.
\newblock \showarticletitle{{Practical evaluation of carrier sensing for a LoRa
  wildlife monitoring network}}. In \bibinfo{booktitle}{\emph{Proc. of 19th
  IFIP Networking Conference}} (Paris, France). \bibinfo{publisher}{IEEE
  Press}, \bibinfo{address}{Piscataway, NJ, USA}, \bibinfo{pages}{10--18}.
\newblock


\bibitem[Oran and Kutscher(2020)]%
        {draft-oran-icnrg-reflexive-forwarding}
\bibfield{author}{\bibinfo{person}{David Oran} {and} \bibinfo{person}{Dirk
  Kutscher}.} \bibinfo{year}{2020}\natexlab{}.
\newblock \bibinfo{booktitle}{\emph{{Reflexive Forwarding for CCNx and NDN
  Protocols}}}.
\newblock \bibinfo{type}{Internet-Draft -- work in progress}~01.
  \bibinfo{institution}{IETF}.
\newblock


\bibitem[Orfanidis et~al\mbox{.}(2019)]%
        {ofjg-cccal-19}
\bibfield{author}{\bibinfo{person}{Charalampos Orfanidis},
  \bibinfo{person}{Laura~Marie Feeney}, \bibinfo{person}{Martin Jacobsson},
  {and} \bibinfo{person}{Per Gunningberg}.} \bibinfo{year}{2019}\natexlab{}.
\newblock \showarticletitle{{Cross-Technology Clear Channel Assessment for
  Low-Power Wide Area Networks}}. In \bibinfo{booktitle}{\emph{16th
  International Conference on Mobile Ad Hoc and Sensor Systems (MASS'19)}}.
  \bibinfo{publisher}{IEEE Computer Society}, \bibinfo{address}{Washington, DC,
  USA}, \bibinfo{pages}{199--207}.
\newblock


\bibitem[Paxson et~al\mbox{.}(2011)]%
        {RFC-6298}
\bibfield{author}{\bibinfo{person}{V. Paxson}, \bibinfo{person}{M. Allman},
  \bibinfo{person}{J. Chu}, {and} \bibinfo{person}{M. Sargent}.}
  \bibinfo{year}{2011}\natexlab{}.
\newblock \bibinfo{booktitle}{\emph{{Computing TCP's Retransmission Timer}}}.
\newblock \bibinfo{type}{RFC} 6298. \bibinfo{institution}{IETF}.
\newblock


\bibitem[Polyzos and Fotiou(2015)]%
        {pf-britu-15}
\bibfield{author}{\bibinfo{person}{George~C. Polyzos} {and}
  \bibinfo{person}{Nikos Fotiou}.} \bibinfo{year}{2015}\natexlab{}.
\newblock \showarticletitle{{Building a reliable Internet of Things using
  Information-Centric Networking}}.
\newblock \bibinfo{journal}{\emph{Journal of Reliable Intelligent
  Environments}} \bibinfo{volume}{1}, \bibinfo{number}{1}
  (\bibinfo{year}{2015}), \bibinfo{pages}{47--58}.
\newblock


\bibitem[Ron et~al\mbox{.}(2020)]%
        {rllcl-paodt-20}
\bibfield{author}{\bibinfo{person}{Dara Ron}, \bibinfo{person}{Chan-Jae Lee},
  \bibinfo{person}{Kisong Lee}, \bibinfo{person}{Hyun-Ho Choi}, {and}
  \bibinfo{person}{Jung-Ryun Lee}.} \bibinfo{year}{2020}\natexlab{}.
\newblock \showarticletitle{{Performance Analysis and Optimization of Downlink
  Transmission in LoRaWAN Class B Mode}}.
\newblock \bibinfo{journal}{\emph{IEEE Internet of Things Journal}}
  \bibinfo{volume}{7}, \bibinfo{number}{8} (\bibinfo{year}{2020}),
  \bibinfo{pages}{7836--7847}.
\newblock


\bibitem[Scott and Burleigh(2007)]%
        {RFC-5050}
\bibfield{author}{\bibinfo{person}{K. Scott} {and} \bibinfo{person}{S.
  Burleigh}.} \bibinfo{year}{2007}\natexlab{}.
\newblock \bibinfo{booktitle}{\emph{{Bundle Protocol Specification}}}.
\newblock \bibinfo{type}{RFC} 5050. \bibinfo{institution}{IETF}.
\newblock


\bibitem[Shang et~al\mbox{.}(2016a)]%
        {saz-dinps-16}
\bibfield{author}{\bibinfo{person}{Wenato Shang}, \bibinfo{person}{Alex
  Afanasyev}, {and} \bibinfo{person}{Lixia Zhang}.}
  \bibinfo{year}{2016}\natexlab{a}.
\newblock \showarticletitle{{The Design and Implementation of the NDN Protocol
  Stack for RIOT-OS}}. In \bibinfo{booktitle}{\emph{Proc. of IEEE GLOBECOM
  2016}}. \bibinfo{publisher}{IEEE}, \bibinfo{address}{Washington, DC, USA},
  \bibinfo{pages}{1--6}.
\newblock


\bibitem[Shang et~al\mbox{.}(2016b)]%
        {sblwy-ndnti-16}
\bibfield{author}{\bibinfo{person}{Wentao Shang}, \bibinfo{person}{Adeola
  Bannis}, \bibinfo{person}{Teng Liang}, \bibinfo{person}{Zhehao Wang},
  \bibinfo{person}{Yingdi Yu}, \bibinfo{person}{Alexander Afanasyev},
  \bibinfo{person}{Jeff Thompson}, \bibinfo{person}{Jeff Burke},
  \bibinfo{person}{Beichuan Zhang}, {and} \bibinfo{person}{Lixia Zhang}.}
  \bibinfo{year}{2016}\natexlab{b}.
\newblock \showarticletitle{{Named Data Networking of Things (Invited Paper)}}.
  In \bibinfo{booktitle}{\emph{Proc. of IEEE International Conf. on
  Internet-of-Things Design and Implementation (IoTDI)}}.
  \bibinfo{publisher}{IEEE Computer Society}, \bibinfo{address}{Los Alamitos,
  CA, USA}, \bibinfo{pages}{117--128}.
\newblock


\bibitem[Siris et~al\mbox{.}(2012)]%
        {svpl-inais-12}
\bibfield{author}{\bibinfo{person}{Vasilios~A. Siris},
  \bibinfo{person}{Christopher~N. Ververidis}, \bibinfo{person}{George~C.
  Polyzos}, {and} \bibinfo{person}{Konstantinos~P. Liolis}.}
  \bibinfo{year}{2012}\natexlab{}.
\newblock \showarticletitle{{Information-Centric Networking (ICN) architectures
  for integration of satellites into the Future Internet}}. In
  \bibinfo{booktitle}{\emph{First AESS European Conference on Satellite
  Telecommunications (ESTEL'12)}}. \bibinfo{publisher}{IEEE},
  \bibinfo{address}{Piscataway, NJ, USA}, \bibinfo{numpages}{6}~pages.
\newblock


\bibitem[Thielemans et~al\mbox{.}(2017)]%
        {tbs-eticl-17}
\bibfield{author}{\bibinfo{person}{Steffen Thielemans}, \bibinfo{person}{Maite
  Bezunartea}, {and} \bibinfo{person}{Kris Steenhaut}.}
  \bibinfo{year}{2017}\natexlab{}.
\newblock \showarticletitle{{Establishing transparent IPv6 communication on
  LoRa based low power wide area networks (LPWANS)}}. In
  \bibinfo{booktitle}{\emph{Wireless Telecommunications Symposium (WTS '17)}}
  (Chicago, IL, USA). \bibinfo{publisher}{IEEE}, \bibinfo{address}{Piscataway,
  NJ, USA}, \bibinfo{pages}{1--6}.
\newblock


\bibitem[Tschudin et~al\mbox{.}(2018)]%
        {ccn-lite}
\bibfield{author}{\bibinfo{person}{Christian Tschudin},
  \bibinfo{person}{Christopher Scherb}, {et~al\mbox{.}}}
  \bibinfo{year}{2018}\natexlab{}.
\newblock \bibinfo{title}{{CCN Lite: Lightweight implementation of the Content
  Centric Networking protocol}}.
\newblock
\newblock
\urldef\tempurl%
\url{http://ccn-lite.net}
\showURL{%
\tempurl}


\bibitem[Vincenzo et~al\mbox{.}(2019)]%
        {vht-idsl-19}
\bibfield{author}{\bibinfo{person}{Valentina~Di Vincenzo},
  \bibinfo{person}{Martin Heusse}, {and} \bibinfo{person}{Bernard
  Tourancheau}.} \bibinfo{year}{2019}\natexlab{}.
\newblock \showarticletitle{{Improving Downlink Scalability in LoRaWAN}}. In
  \bibinfo{booktitle}{\emph{IIEEE International Conference on Communications
  (ICC'19)}}. \bibinfo{publisher}{IEEE}, \bibinfo{address}{Piscataway, NJ,
  USA}, \bibinfo{numpages}{7}~pages.
\newblock


\bibitem[Yapar et~al\mbox{.}(2019)]%
        {yte-talsa-19}
\bibfield{author}{\bibinfo{person}{Gokcer Yapar}, \bibinfo{person}{Tuna Tugcu},
  {and} \bibinfo{person}{Orhan Ermis}.} \bibinfo{year}{2019}\natexlab{}.
\newblock \showarticletitle{{Time-Slotted ALOHA-based LoRaWAN Scheduling with
  Aggregated Acknowledgement Approach}}. In \bibinfo{booktitle}{\emph{25th
  Conference of Open Innovations Association (FRUCT'19)}}.
  \bibinfo{publisher}{IEEE}, \bibinfo{address}{Piscataway, NJ, USA},
  \bibinfo{pages}{383--390}.
\newblock


\bibitem[Yu et~al\mbox{.}(2021)]%
        {yzmmz-panbs-21}
\bibfield{author}{\bibinfo{person}{Tianyuan Yu}, \bibinfo{person}{Zhiyi Zhang},
  \bibinfo{person}{Xinyu Ma}, \bibinfo{person}{Philipp Moll}, {and}
  \bibinfo{person}{Lixia Zhang}.} \bibinfo{year}{2021}\natexlab{}.
\newblock \bibinfo{booktitle}{\emph{{A Pub/Sub API for NDN-Lite with Built-in
  Security}}}.
\newblock \bibinfo{type}{Technical Report} NDN-0071.
  \bibinfo{institution}{NDN}.
\newblock


\bibitem[Zhang et~al\mbox{.}(2014)]%
        {zabjc-ndn-14}
\bibfield{author}{\bibinfo{person}{Lixia Zhang}, \bibinfo{person}{Alexander
  Afanasyev}, \bibinfo{person}{Jeffrey Burke}, \bibinfo{person}{Van Jacobson},
  \bibinfo{person}{kc claffy}, \bibinfo{person}{Patrick Crowley},
  \bibinfo{person}{Christos Papadopoulos}, \bibinfo{person}{Lan Wang}, {and}
  \bibinfo{person}{Beichuan Zhang}.} \bibinfo{year}{2014}\natexlab{}.
\newblock \showarticletitle{{Named Data Networking}}.
\newblock \bibinfo{journal}{\emph{SIGCOMM Comput. Commun. Rev.}}
  \bibinfo{volume}{44}, \bibinfo{number}{3} (\bibinfo{year}{2014}),
  \bibinfo{pages}{66--73}.
\newblock


\bibitem[Zorbas et~al\mbox{.}(2020)]%
        {zakp-ttlii-20}
\bibfield{author}{\bibinfo{person}{Dimitrios Zorbas}, \bibinfo{person}{Khaled
  Abdelfadeel}, \bibinfo{person}{Panayiotis Kotzanikolaou}, {and}
  \bibinfo{person}{Dirk Pesch}.} \bibinfo{year}{2020}\natexlab{}.
\newblock \showarticletitle{{TS-LoRa: Time-slotted LoRaWAN for the Industrial
  Internet of Things}}.
\newblock \bibinfo{journal}{\emph{Computer Communications}}
  \bibinfo{volume}{153} (\bibinfo{year}{2020}), \bibinfo{pages}{1--10}.
\newblock


\end{thebibliography}

\label{lastpage}
\end{document}